\documentclass[12pt]{article}

\usepackage{epsf,latexsym}

\begin{document}
  \title{Static and Dynamic Monte Carlo Simulations of Br Electrodeposition on Ag(100)
}
\author{
\centerline{S. J. Mitchell, G. Brown, P. A. Rikvold\footnote{Corresponding author.
E-mail:  rikvold@csit.fsu.edu  FAX:  1-850-644-0098}}\\
\centerline{\footnotesize\it Center for Materials Research and Technology,}\\
\centerline{\footnotesize\it Department of Physics, and}\\
\centerline{\footnotesize\it School of Computational Science and Information Technology,}\\
\centerline{\footnotesize\it Florida State University, 
Tallahassee, Florida 32306-4351, USA}\\
}
\date{\today}
\maketitle

\begin{abstract}
We study the static and dynamic properties of bromine electrosorption
onto single-crystal silver (100) electrodes by Monte Carlo simulation.
At room temperature the system displays a second-order phase transition
between a low-coverage disordered phase at more negative electrode potentials
and a c$(2 \times 2)$ ordered phase with bromine coverage $1/2$ at more positive potentials.
We explore the phase diagram and demonstrate that the broad shoulder 
observed in room-temperature cyclic voltammograms
is due to local fluctuations resembling ordered phases with coverage $1/4$,
which exist in the model at much lower temperatures.
We construct a dynamic Monte Carlo algorithm using a thermally activated
stochastic barrier-hopping model for the microscopic dynamics.
We use this algorithm to study the phase ordering and disordering processes
following sudden potential steps between the disordered phase and the c$(2 \times 2)$ phase,
and to study the sweep-rate dependence in simulated cyclic-voltammetry experiments.
\end{abstract}


\section{Introduction}
\label{sec:I}

A number of experimental techniques have emerged in recent years
which enable studies of the structure and dynamics of
electrode-electrolyte interfaces {\it in situ}, 
i.e., without removing the electrode from the liquid \cite{WIEC99}.
This new atomic scale information allows greater understanding
of the deposition process where, 
in contrast to high-vacuum surface-science methods, 
the electrochemical environment provides direct experimental
control over the electrochemical potentials of the adsorbates,
rather than over their coverages or fluxes.  
These new {\it in situ} experimental techniques, 
together with new methods to manipulate adsorbates at the interface on the nanometer scale \cite{KOLB98},
promise to make electrodeposition an even more important method for low-cost,
high-quality materials synthesis than it is today \cite{GREG91}.  
These developments emphasize the importance of developing theoretical and computational methods to study
the dynamical aspects of electrodeposition in microscopic detail.

Motivated by these developments in electrochemical surface science, 
we here present a static and dynamic Monte Carlo (MC) simulation study of a simple system 
which can serve as a prototype for more complicated ones of applied interest.
This system is a submonolayer of Br electrodeposited onto a single-crystal Ag(100) electrode.
Recently it was investigated at room temperature
by {\it in-situ\/} surface X-ray scattering (SXS) and chronocoulometry by Ocko, Wang, and 
Wandlowski \cite{OCKO:BR/AG,WANDLOWSKI:HALIDE00}. 
That study detected a second-order phase transition between a disordered low-coverage phase 
at more negative electrode potentials and a doubly degenerate 
c$(2 \times 2)$ ordered phase with coverage $1/2$ at more positive potentials.
(This phase has also been observed by low-energy electron scattering (LEED)
for Br adsorbed on Ag(100) in vacuum \cite{TAYL85}.)
The transition was shown to belong to the two-dimensional Ising universality class \cite{OCKO:BR/AG}.
The scattering data \cite{OCKO:BR/AG,WANDLOWSKI:HALIDE00} also indicate that Br adsorbs at the four-fold hollow
sites between the surface Ag atoms, consistent with recent
{\it in situ\/} X-ray absorption fine structure (XAFS) results \cite{ENDO99}.
The fluctuations around the adsorption sites are small,
suggesting that a lattice-gas approximation for the adlayer
structure and dynamics is appropriate.
Here we investigate in detail such a lattice-gas model with a nearest-neighbor
excluded-volume interaction and lateral dipole-dipole repulsive interactions 
\cite{KOPER:HALIDE}.
The dipole interaction strength and the electrosorption valency,
which are parameters of the model, are estimated by fitting 
simulated equilibrium adsorption isotherms to experimental chronocoulometric
data \cite{WANG:FRUM}. 

Cyclic voltammograms (CVs) at room temperature show two main features 
\cite{OCKO:BR/AG,WANDLOWSKI:HALIDE00,ENDO99,WANG:FRUM,VALETTE:AG,OCKO97B}:
a broad shoulder at more negative potentials 
which ends in a sharp peak at more positive potentials.
By SXS it was confirmed that the sharp peak corresponds 
to a second-order phase transition in the layer of adsorbed Br 
\cite{OCKO:BR/AG}.  However, the cause 
of the broad shoulder has been less clear. 
It was previously suggested \cite{VALETTE:AG} that interactions with surface water are 
responsible for the shoulder, 
but equilibrium MC simulations by Koper \cite{KOPER:HALIDE,KOPE98C}, as well as our own work,
show the same features in simulations without water.
We show here that the broad shoulder in the simulated CV is due to fluctuations 
within the disordered adlayer.
Locally, these fluctuations resemble other ordered phases with coverage 1/4, 
which exist in the model at much lower temperatures.

Going beyond these equilibrium simulations, we further
present a dynamic MC study
of the ordering and disordering processes between the disordered phase 
and the c$(2 \times 2)$ phase,
assuming a perfectly stirred solution so that bulk diffusion effects can be ignored.
For ordering, we expect similar behavior to the random sequential adsorption
process of the hard-square model with lateral diffusion \cite{WANG93,EISE98}.
At late stages of the ordering, this dynamic model is expected to be in the same
dynamic universality class as the ferromagnetic \cite{GUNTON:DYNTRAN}
or antiferromagnetic \cite{SAHN81} Ising model in zero magnetic
field following sudden cooling from infinite temperature 
to a temperature below the critical point. 
These phase ordering dynamics with non-conserved order parameter,
often called Model A in the literature on dynamics of phase ordering 
and phase separation \cite{GUNTON:DYNTRAN,HOHE77},
are driven by the reduction of interface curvature \cite{LIFS62,ALLE79}.
We further present dynamic results for the disordering process
and discuss how simulation results for this process can
be compared with experiment.

We also apply the dynamic MC algorithm to study sweep-rate dependence in simulated CV experiments.
Results are presented for the response of the system to a wide range of sweep rates,
and we suggest some comparisons between simulation results and experiment.
Some preliminary results of this study were included in Refs.\ \cite{ME:JEAC00,ME:UGA00}.

The remainder of this paper is organized as follows.
In Sec.~\ref{sec:LGM} 
we introduce the lattice-gas model and the relation between the 
electrochemical potential and the experimentally measured electrolyte concentration and electrode potential.
In Sec.~\ref{sec:eqMC} we present our equilibrium results, 
and in Sec.~\ref{sec:dynMC} we introduce our dynamic MC algorithm.
In Sec.~\ref{sec:step} 
we present dynamic simulations of sudden potential-step experiments, 
and in Sec.~\ref{sec:CV} we present dynamic simulations of CV experiments.
In Sec.~\ref{sec:D} we give a summary and our conclusions.
The Appendix describes the nonlinear fitting procedure used to determine the 
lattice-gas parameters.
Animations and additional figures are available on the World Wide Web \cite{WWW}.

\section{Lattice-Gas Model}
\label{sec:LGM}

The lattice-gas model consists of a two-dimensional $L\times L$ square array 
of Br adsorption sites, 
corresponding to the four-fold hollow sites on the Ag(100) surface.
Periodic boundary 
conditions were used to eliminate edge effects in the relatively small
systems it is possible to simulate.
The grand-canonical lattice-gas Hamiltonian is
\begin{equation}
{\mathcal H}=-\sum_{i<j} \phi_{ij} c_i c_j 
- \overline{\mu} \sum_{i=1}^{L^2} c_i 
\;,
\label{eq:hamil}
\end{equation}
where $i$ and $j$ denote sites on the lattice,
$c_i$ denotes the occupation state of site $i$,
which is either 0 (empty) or 1 (occupied),
$\sum_{i<j}$ denotes a sum over all pairs of sites on the lattice, 
$\phi_{ij}$ denotes the lateral interaction energy of the pair $(i,j)$,
and $\overline{\mu}$ is the electrochemical potential.
The sign convention used here is such 
that $\phi_{ij}<0$ denotes a repulsive interaction,
and $\overline{\mu}>0$ favors adsorption.
We measure $\phi_{ij}$ and 
$\overline{\mu}$ in units of meV/pair and meV/particle, respectively.
[For brevity we write both simply as meV.]

The electrochemical potential $\overline{\mu}$ is the intensive variable
which controls the average number of Br in the adlayer.
In the weak-solution approximation it is related to the electrode potential by 
\begin{equation}
\overline{\mu} = \overline{\mu}_0 
+ k_{\rm B} T \ln \frac{[C]}{[C_{0}]} - e\gamma E 
\;,
\end{equation}
where $\overline{\mu}_0$ is an arbitrary reference level,
$k_{\rm B}$ is Boltzmann's constant,
$T$ is the absolute temperature,
$e$ is the elementary charge unit,
$[C]$ is the concentration of Br$^-$ in solution, 
$[C_{0}]$ is an arbitrary reference concentration (here we use $C_0=1$~mM),
$\gamma$ is the electrosorption valency
($\gamma e$ is the charge transfered through the external circuit
for each Br ion adsorbed on the electrode \cite{VETTER1,VETTER2,SCHM96}), 
and $E$ is the electrode potential in mV.

A variety of lateral 
interactions have been previously explored for this model, 
including finite and infinite nearest-neighbor repulsion 
and screened dipole-dipole interactions \cite{KOPER:HALIDE}.
From these previous studies and our own exploratory simulations, 
we find that the 
interactions are quite adequately described with a nearest-neighbor
excluded-volume interaction 
(infinite nearest-neighbor repulsion) related to the large ionic radius of Br$^-$ (1.96~{\AA} \cite{CRC99}),
plus a long-range dipole-dipole repulsion.
The pairwise lateral interactions used here are
\begin{equation}
\phi (r)= 
\left\{
\begin{array}{ll}
-\infty & r=1 \\
\frac{ 2^{3/2} \phi_{\rm nnn}}{r^3} & \sqrt{2} \le r \le 5 \\
0 & r > 5
\end{array} 
\right. \; ,
\end{equation}
where $r$ is the separation of an interacting Br pair within the adlayer,
measured in units of 
the Ag(100) lattice spacing ($a=2.889$~{\AA} \cite{OCKO:BR/AG}),
and $\phi_{\rm nnn}$ (which is negative) is the lateral dipole-dipole repulsion 
between next-nearest neighbors. 
In other work it has been noticed that reducing the infinite nearest-neighbor 
repulsion to finite but large values has very little effect on the isotherms \cite{KOPER:HALIDE}.
All interactions are truncated beyond five lattice spacings to simplify computation.
This truncation can be seen as an approximate cut-off due to screening.
With this cut-off, the total interaction energy of a fully ordered c$(2\times 2)$
phase is approximately $87\%$ of the interaction energy without a cut-off.

The Ag(100) substrate provides a square adsorption lattice of hollow sites for
the commensurate Br adlayer.
The nearest-neighbor excluded-volume interactions give rise to two different sublattices
which have a geometry corresponding to the two colors on a checkerboard.
Each of the two degenerate c$(2 \times 2)$ phases can occupy 
only one of the two sublattices, which we label $A$ and $B$.
The $A$ sublattice coverage is the fraction of occupied sites on sublattice $A$,
defined as $\Theta_A = N_A^{-1} \sum_{i\in A}^{N_A} c_{i}$,
where $N_A$ is the number of sites on sublattice $A$
and $\sum_{i\in A}^{N_A}$ denotes a sum over all sites on that sublattice.
The $B$ sublattice coverage, $\Theta_B$, is analogously defined.
For this system, there are two linearly independent observables of interest:
the total Br coverage, $\Theta = (\Theta_A+\Theta_B)/2$,
and the staggered coverage, $\Theta_{\rm S} = \Theta_A - \Theta_B$,
which is the order parameter for the c$(2 \times 2)$ phase.

\section{Equilibrium Monte Carlo Simulations}
\label{sec:eqMC}

Monte Carlo simulations were performed at room temperature ($k_{\rm B} T=25$~meV, 
corresponding to $T\approx290$~K) to determine the equilibrium 
properties of the Br adlayer for different values of $\overline{\mu}$.
For these equilibrium simulations, we used the Metropolis algorithm \cite{BINDER:MCBOOK}
with single-site updates at randomly selected sites.

The main purpose of these equilibrium simulations was to obtain 
adsorption isotherms for parameter estimation by fitting to experimental data. 
The most important region of comparison is in the disordered phase, 
where the fluctuations are small. 
It was therefore sufficient to use relatively small systems with $L=$32.
No significant differences were noticed between data for $L=32$ and $L=128$ (see below).
No systematic finite-size scaling was performed near the phase transition. 
Thus, the results obtained in the critical region are only approximate.

For a given temperature and chemical potential, we generated a set of samples
from which we computed the averages $\langle \Theta(T,\overline{\mu}) \rangle$ 
and $\langle | \Theta_{\rm S}(T,\overline{\mu}) | \rangle$, 
where $\langle \cdot \rangle$ denotes an average over a set 
of statistically independent equilibrium samples.
For convenience, we drop the $\langle \cdot \rangle$ notation for the equilibrium isotherms.

The critical value of the electrochemical potential and the shape of the 
adsorption isotherm in the disordered phase are quite sensitive to the strength
of the long-range repulsive interactions. 
To determine the dipole interaction strength, $\phi_{\rm nnn}$,
and the electrosorption valency, $\gamma$,
we therefore simultaneously compared the simulated room-temperature adsorption isotherm, $\Theta(\bar{\mu})$,
with experimental isotherms, $\Theta^{\rm exp}(E)$, obtained from chronocoulometry \cite{OCKO:BR/AG}.
The nonlinear fitting method is described in the Appendix.
The resulting parameter values are $\phi_{\rm nnn}=-26\pm 2$~meV, $\gamma=-0.73\pm 0.03$,
and $\bar{\mu}_0=0.67\pm 0.03$~meV. 
These results are consistent with those found previously by others \cite{WANDLOWSKI:HALIDE00,KOPER:HALIDE,WANG:FRUM} 

The fitted MC isotherm is shown together with 
the chronocoulometric coverage isotherms in Fig.~\ref{fig:exp}. 
For the two lowest concentrations, 
the agreement is very good over the whole range of electrode potentials, 
while the agreement in the ordered phase is somewhat less satisfactory for the highest concentration.
This disagreement at higher concentrations suggests a break-down 
in the weak-solution approximation and the ultimate need for a more detailed treatment of the solution.

\begin{figure}
{\epsfxsize=6in \epsfbox{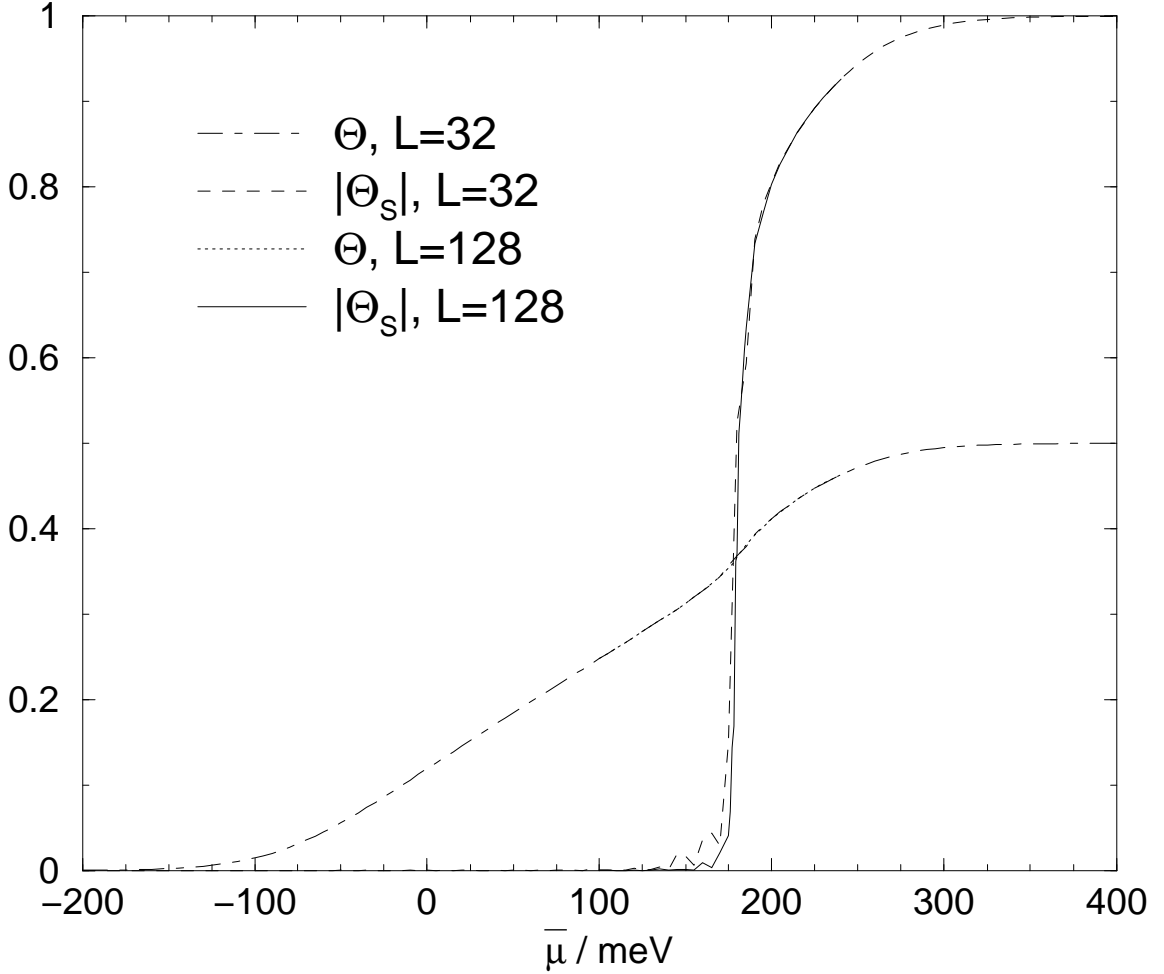}}
\caption[]{
MC isotherms for $L=32$ and $L=128$, $T\approx 290$~K, and $\phi_{\rm nnn}=-26$~meV.
A second-order phase transition between a low-coverage disordered phase and a c$(2 \times 2)$
phase with $\Theta=1/2$ is observed at $\bar{\mu}_c \approx 180$~meV.
$\Theta_{\rm S}$ is the order parameter for the c$(2 \times 2)$ phase.
In an infinite system, $|\Theta_{\rm S}| \propto (\bar{\mu}-\bar{\mu}_c)^{1/8}$
for $\bar{\mu}\ge\bar{\mu}_c$.
The $L=32$ isotherms are generated from 
10,000 independent samples for each value of $\bar{\mu}$.
For $L=128$, the number of independent samples per $\bar{\mu}$ value 
varies between 1000 (away from the transition) and 3000 (near the transition).
Thus, the statistical errors are larger than for $L=32$.
}
\label{fig:isoth}
\end{figure}

Figure~\ref{fig:isoth} shows the simulated $\Theta(\bar{\mu})$ and $|\Theta_{\rm S}(\bar{\mu})|$
at $T\approx 290$~K, with $\phi_{\rm nnn}=-26$~meV as obtained from the fits 
to the experimental isotherms. 
The order parameter, $\Theta_{\rm S}$, is zero in the disordered phase, 
while it is near $\pm 1$ deep in either of the two degenerate ordered phases.
A second-order phase transition between the low-coverage disordered phase
and the c$(2 \times 2)$ ordered phase is observed at a critical electrochemical 
potential, $\bar{\mu}_c \approx 180\pm5$~meV.
As $\bar{\mu} \rightarrow \bar{\mu}_c^+$, 
$|\Theta_{\rm S}|\propto (\bar{\mu}-\bar{\mu}_c)^{1/8}$ \cite{OCKO:BR/AG},
as expected for the Ising universality class. 

The coverage isotherm shows a long, gradual slope in the disordered phase, 
which is not present in a model with only nearest-neighbor exclusion
(called the hard-square (HS) model) \cite{BAXTER:HS,RACZ80}.
As a result, $\bar{\mu}_c$ is considerably larger than that of the pure hard-square model, 
$\bar{\mu}_c^{\rm HS} \approx 33$~meV
\cite{BAXTER:HS,RACZ80}\footnote{
The value $\bar{\mu}_c^{\rm HS} \approx 33$~meV is obtained as
$\bar{\mu}_c^{\rm HS} = k_{\rm B} T \ln z_c$, using precise numerical values
for the critical activity of the hard-square model, $z_c = 3.7962(1)$
\protect\cite{BAXTER:HS} and $z_c = 3.7959(1)$ \protect\cite{RACZ80},
obtained by series expansions and transfer-matrix finite-size scaling, respectively.}.
The simulated and experimental isotherms show coverages at $\bar{\mu}_c$
of about $0.37\pm 0.01$ and $0.36\pm 0.02$, respectively.
Within statistical errors, this is the same result as the pure hard-square 
model, $\Theta_c^{\rm HS} = 0.368$
\cite{RACZ80,REE66}\footnote{
Precise numerical results have been obtained for the pure hard-square model
by transfer-matrix calculations, including $\Theta_c^{\rm HS} = 0.36776(1)$
\protect\cite{REE66} and $\Theta_c^{\rm HS} = 0.368(1)$ \protect\cite{RACZ80}.}.
The adsorption isotherm, $\Theta(\bar{\mu})$, has its maximum slope at $\bar{\mu}_c$.
Corresponding maxima in the slopes are also seen in the experimental isotherms in Fig.~\ref{fig:exp}.
Simulation data are included in Fig.~\ref{fig:isoth} for both $L=32$ and $L=128$.
On the scale of this figure, differences between the two sizes are visible only in the data
for $|\Theta_{\rm S}|$ very close to $\bar{\mu}_c$,
where ${\rm d}|\Theta_{\rm S}|/{\rm d}\bar{\mu}$ is expected to diverge with increasing $L$
as $L^{7/4}$ 
\cite{BIND92C}\footnote{
At $\bar{\mu}_c$ in a finite system of size $L$,
${\rm d}|\Theta_{\rm S}|/{\rm d}\bar{\mu} \propto L^{\gamma/\nu}$, 
where $\gamma$ and $\nu$ are the critical exponents for the ordering susceptibility and the correlation length, respectively.
For the two-dimensional Ising model $\gamma=7/4$ and $\nu=1$ \cite{BINDER:MCBOOK,BIND92C}.}.

A quasi-equilibrium CV (the CV in the limit of vanishing potential sweep rate)
can be obtained from the simulated equilibrium isotherm using the simple relation, 
\begin{equation}
{\mathcal J}=-\gamma e M \frac{{\rm d}\Theta}{{\rm d}t}=
-\gamma e M \frac{{\rm d}\Theta}{{\rm d}\bar{\mu}}\frac{{\rm d}\bar{\mu}}{{\rm d}E}\frac{{\rm d}E}{{\rm d}t}=
\gamma^2 e^2 M\frac{{\rm d}\Theta}{{\rm d}\bar{\mu}}\frac{{\rm d}E}{{\rm d}t}
\propto \frac{{\rm d}\Theta}{{\rm d}\bar{\mu}} \;,
\end{equation}
where ${\mathcal J}$ is the voltammetric 
current density (oxidation currents positive), $M$ is the total number of 
adsorption sites per unit area 
($a^{-2} = 1.198 \times 10^{15}$ sites/cm$^2$ \cite{OCKO:BR/AG}), 
and ${{\rm d}E}/{{\rm d}t}$ is the sweep rate.
Figure~\ref{fig:qCV} shows ${{\rm d}\Theta}/{{\rm d}\bar{\mu}}$, 
which exhibits the same broad shoulder and sharp peak
seen in experimental CVs \cite{OCKO:BR/AG,ENDO99,VALETTE:AG}.
The shoulder has previously been attributed 
to interactions with surface water \cite{VALETTE:AG}.
However, the fact that the lattice-gas model 
does not contain explicit water molecules indicates that the 
shoulder is most likely caused by configurational fluctuations in the Br adlayer,
as was also pointed out by Koper \cite{KOPER:HALIDE}.
Differences between the simulation data for $L=32$ and 128 are only visible near the peak,
which is expected to diverge logarithmically with $L$ 
\cite{RIKVOLD85}\footnote{
At $\bar{\mu}_c$ ${\rm d}\Theta/{\rm d}\bar{\mu}$ is expected to be proportional to the heat capacity of the adlayer.
The heat capacity of the two-dimensional Ising model diverges logarithmically with $L$ 
\cite{BINDER:MCBOOK,BIND92C,RIKVOLD85}.}.
This quasi-equilibrium CV is compared with dynamically simulated CVs for
different potential sweep rates in Sec.~\ref{sec:CV}.

\begin{figure}
{\epsfxsize=6in \epsfbox{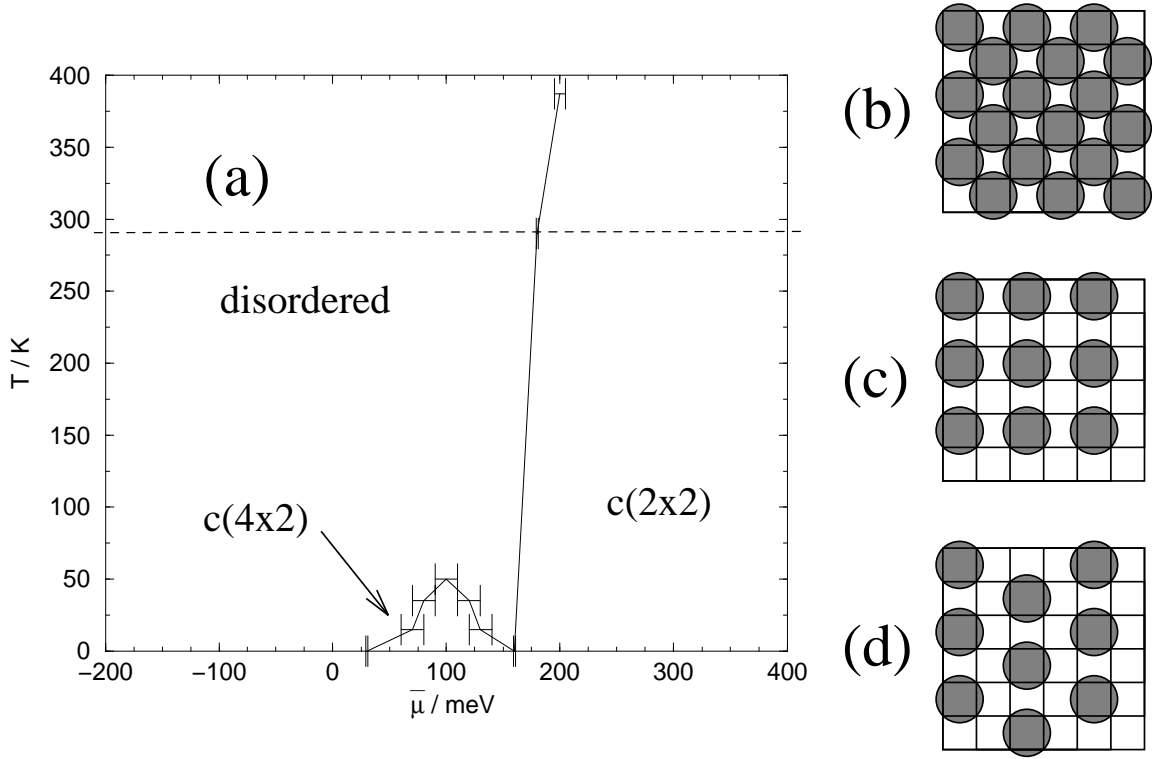}}
\caption[]{
Approximate phase 
diagram and observed phases for Br electrodeposited onto Ag(100).
(a): Approximate phase diagram 
obtained with a two-site heat-bath MC algorithm for $L=16$.
The horizontal dashed line indicates room temperature.
The disordered phase and both stable ordered phases are labeled.
The three observed ordered phases are 
(along with symmetry equivalents)
(b): c$(2 \times 2)$ with $\Theta=1/2$ (stable),
(c): p$(2 \times 2)$ with $\Theta=1/4$ (metastable),  
(d): c$(4 \times 2)$ with $\Theta=1/4$ (stable).
}
\label{fig:PD}
\end{figure}

Having obtained an estimate for the lateral interaction constant  
by equilibrium simulations at room temperature,
we briefly explore the full phase diagram of the model with this interaction strength,
shown in Fig.~\ref{fig:PD}. 
We expect the phase diagram for this system to be topologically 
similar to that for the hard-square model with next-nearest neighbor
repulsion \cite{KINZEL:HSNNN} (even though the low-coverage ordered
phases in the two systems are different). 
In addition to the disordered phase and the two equilibrium ordered phases 
[the low-temperature c$(4 \times 2)$ phase with $\Theta=1/4$ 
and the c$(2 \times 2)$ phase with $\Theta=1/2$ which is observed at all temperatures for 
sufficiently large $\bar{\mu}$], there is also a metastable phase, p$(2 \times 2)$ with $\Theta=1/4$,
which lies closely above the c$(4 \times 2)$ phase in energy.
To facilitate equilibration at low temperatures, in this simulation
we used a heat-bath algorithm which updates clusters \cite{RIKV95} consisting of 
randomly chosen nearest-neighbor pairs of sites.
The presence of the p$(2 \times 2)$ phase made low-temperature 
equilibration very difficult, even with the pair-update algorithm.
Other phases with $\Theta<1/4$ probably also exist as a consequence of the long-range repulsion;
however, we expect these phases to be present only at extremely low temperatures,
and their effects should be negligible at room temperature.
The phase boundaries have been determined by ground-state calculations for $T=0$,
by the location of the sharp peak in the quasi-equilibrium CV for $T>100$~K,
and by inspection of the pair-update MC sample configurations for $0<T<100$~K.

The broad shoulder in the CV is caused by
configurational fluctuations in the disordered phase,
which locally resemble the $\Theta=1/4$ phases \cite{RIKV91A,RIKV91B}.
At the experimentally unrealistic low temperatures where the c$(4 \times 2)$ phase is observed in the model,
the shoulder breaks up into two transitions between the disordered phase and the c$(4 \times 2)$ phase.
Although the c$(4 \times 2)$ phase is not experimentally observable in this system, 
local fluctuations which resemble the $\Theta=1/4$ phases can be seen in the model in 
equilibrium at room temperature and should be observable in the diffuse intensity of SXS experiments. 
We therefore next present simulated equilibrium SXS intensities in the 
room-temperature disordered phase at $\Theta \approx 1/4$. 

\begin{figure}
{\epsfxsize=6in \epsfbox{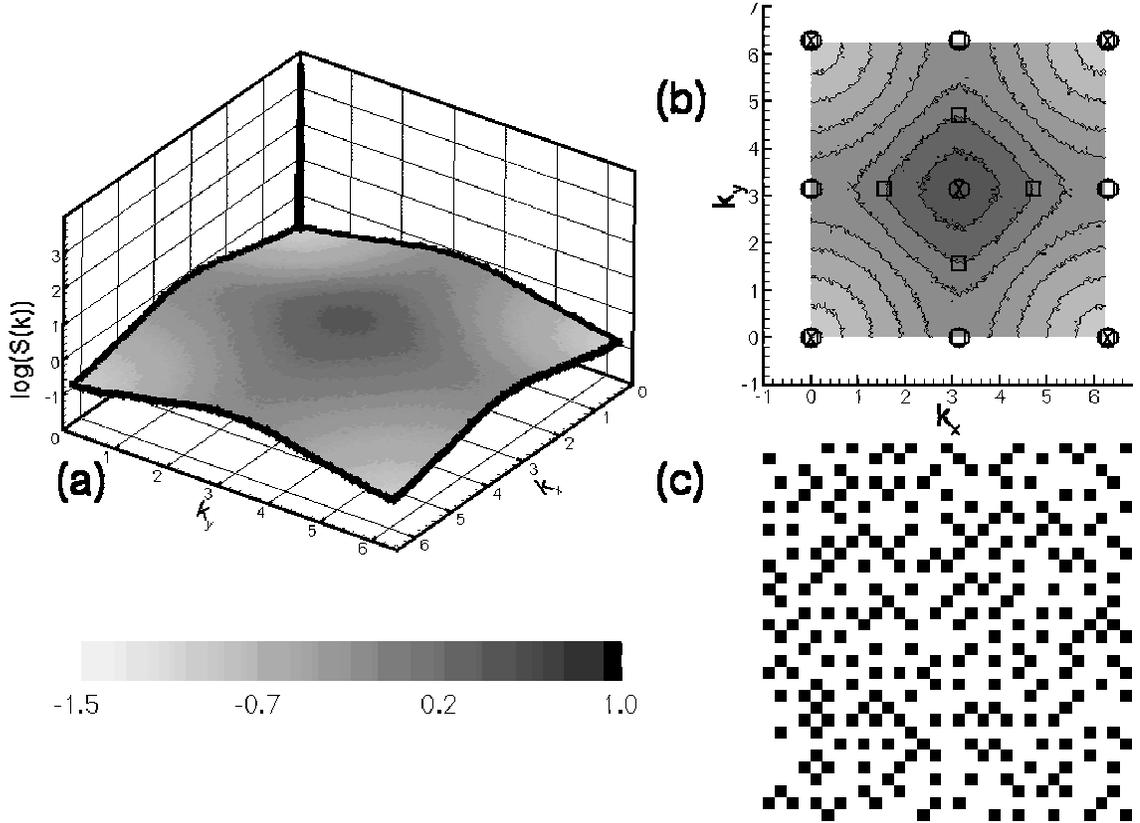}}
\caption[]{
The structure factor, $S(\vec{k})$, 
which is proportional to the surface X-ray scattering (SXS) intensity. Equilibrium 
simulation at room temperature for $L=256$ and $\bar{\mu}=+100$~meV,
averaged over 1,000 independent configurations.
(a): Three-dimensional view of $S(\vec{k})$, 
with the grayscale indicating $\log_{10}[S(\vec{k})]$.
The delta-function peak in the far corner is the Bragg peak at (0,0),
corresponding to $\Theta \approx 1/4$.
(b): Contour plot of $S(\vec{k})$ with the same grayscale as in (a). 
The symbols show where Bragg peaks would occur for fully ordered phases
[c$(2 \times 2)$, $\times$; p$(2 \times 2)$, $\bigcirc$; c$(4 \times 2)$, $\Box$]. 
The peaks on the zone boundary ($k_x=2\pi$, $k_y=2\pi$) are also shown for clarity.
(c): A $32 \times 32$ portion of a real-space configuration.
Small domains resembling all three ordered phases 
shown in Fig.~\protect\ref{fig:PD} are noticeable.
}
\label{fig:eqS}
\end{figure}

We define the structure factor as $S(\vec{k})=|\hat{c}({\vec k})|^2$, 
where $|\hat{c}({\vec k})|^2$ denotes the squared absolute value of the complex Fourier
transform of the real-space adlayer configuration $\{c_i\}$,
and $\vec{k}=(k_x,k_y)$ is the wave vector in the plane of the adlayer
(given in units of the inverse lattice constant, $a^{-1}$).
Equivalently, $S(\vec{k})$ can be expressed as the Fourier transform of
the adsorbate correlation function, $C(\vec{r_i} - \vec{r_j}) = \langle c_i c_j \rangle$. 
The Fourier transform is normalized such that $S(\vec{k}=(0,0))=(L \Theta)^2$.
In the Born approximation, the structure factor is proportional to the SXS intensity \cite{YANG}.
In Fig.~\ref{fig:eqS}(a) and (b) we show the simulated 
structure factor for the disordered room-temperature phase at 
$\bar{\mu} =+100$~meV, where $\Theta \approx 1/4$. Small domains of 
c$(2 \times 2)$, c$(4 \times 2)$, and p$(2 \times 2)$ symmetry are noticeable
in snapshots of the adsorbate configurations [Fig.~\ref{fig:eqS}(c)], 
and they also affect the simulated diffuse scattering intensity. 
Referring to Fig.~\ref{fig:eqS}(b), Bragg peaks would occur at the points marked
$\times$ for a fully ordered $c(2\times 2)$ phase,
at the points marked $\bigcirc$ for $p(2\times 2)$,
and at the points marked $\Box$ for $c(4\times 2)$.
Due to the disorder, all of these peaks [except for the one at $\vec{k}=(0,0)$] are very broad 
and combine to give the anisotropic diffuse scattering intensity shown in the figure.
The individual peaks would become narrower and more distinct
at temperatures less than room temperature (but still above the critical temperature
for the low-temperature ordered phase) or with stronger dipole-dipole interactions.

\section{Dynamic Monte Carlo Simulations}
\label{sec:dynMC}

Next we investigate the dynamics of the Br adlayer using a dynamic MC algorithm.
The microscopic processes considered are adsorption/desorption, in which a particle enters or leaves
the adlayer, as well as nearest-neighbor and next-nearest neighbor lateral diffusion.
The thermally activated, stochastic barrier-hopping picture of the microscopic dynamics used here
is described in further detail elsewhere \cite{ME:ECCHAPT}.

\begin{figure}
{\epsfxsize=6in \epsfbox{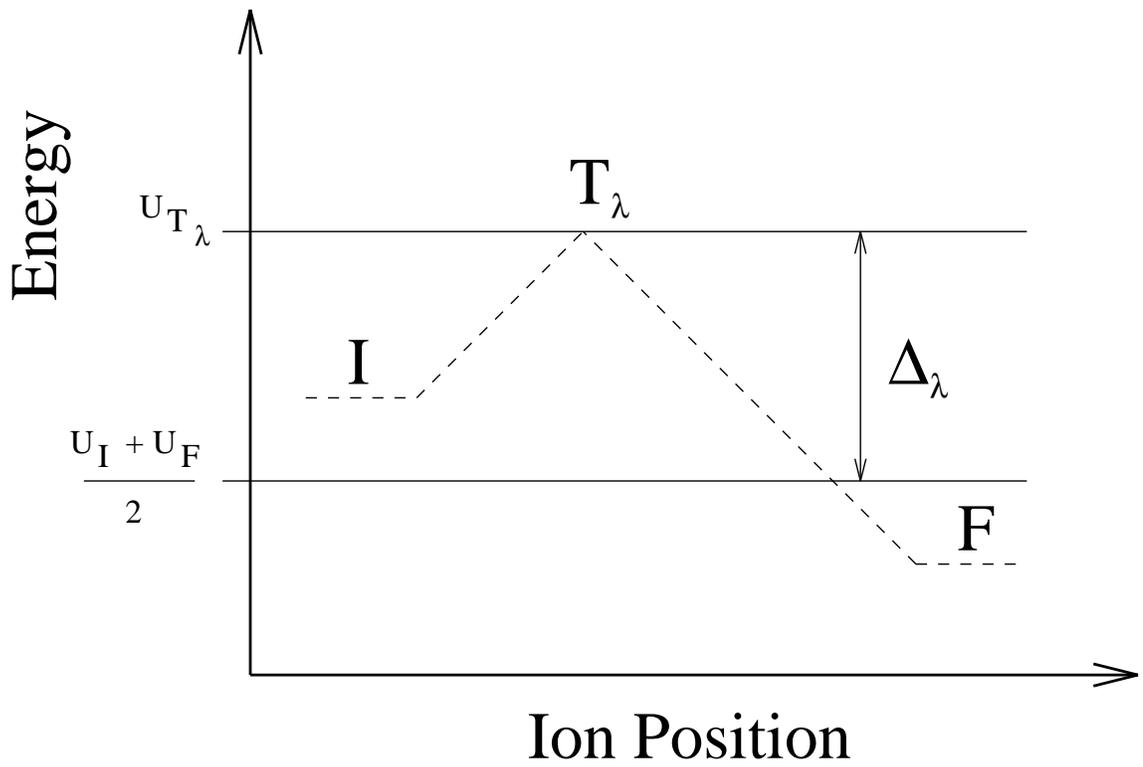}}
\caption[]{
Schematic energy landscape within the Butler-Volmer approximation.
The states $I$ and $F$ denote lattice-gas states.
The state $T_\lambda$ denotes the intermediate transition state
for a microscopic process of type $\lambda$.
The ``bare'' barrier associated with $\lambda$ is denoted $\Delta_\lambda$.
}
\label{fig:barr}
\end{figure}

We assume a simple linear form for the local energy landscape, 
which is equivalent to the Butler-Volmer approximation
with symmetry parameter 1/2 \cite{ME:ECCHAPT,BARD80}.
See Fig.~\ref{fig:barr}.
The initial and final states, $I$ and $F$, respectively,
are represented by lattice-gas configurations and differ by 
a single microscopic move. We label each type of microscopic 
move (adsorption, desorption, nearest-neighbor
and next-nearest neighbor diffusion) by the index $\lambda$.
The state $T_\lambda$ represents 
an intermediate transition state of higher energy between $I$ and $F$,
which is associated with the move $\lambda$.
This intermediate state cannot be represented by a lattice-gas configuration,
and we associate with it a ``bare'' barrier $\Delta_\lambda$.
The energy of the transition state is then approximated as\\
\begin{equation}
U_{T_\lambda}=\frac{U_I+U_F}{2}+\Delta_\lambda \;,
\end{equation}
where $U_{T_\lambda}$, $U_I$, and $U_F$ are the energies 
of the states $T_\lambda$, $I$, and $F$, respectively.

Although other choices of the transition probability are used in the literature as well \cite{TAN:TDA},
we here approximate the probability per MC step per site (MCSS) of making a transition from $I$ to $F$
by the simple one-step Arrhenius rate \cite{ME:ECCHAPT,KANG89,UEBI97}
\begin{equation}
R(F|I)=\nu \exp{\left(-\frac{U_{T_\lambda}-U_I} {k_{\rm B} T}\right)}
=
\nu \exp{\left(-\frac{\Delta_\lambda}{k_{\rm B} T}\right)}
\exp{\left(-\frac{U_F-U_I}{2 k_{\rm B} T}\right)} \;,
\label{eq:dynrate}
\end{equation}
where the rate constant $\nu$ (in units of (MCSS)$^{-1}$)
sets the overall time scale for the simulation and is assumed to be the same for all microscopic processes.

Dynamic Monte Carlo steps proceed in the following way.
First, a lattice site $j$ is randomly selected. 
If $j$ is empty, only adsorption may be attempted.
If $j$ is occupied, both desorption and lateral diffusion may be attempted.
In general, some moves will be excluded by the presence of other 
adatoms (corresponding to $U_F=+\infty$).
Next, a weighted list is constructed of new potential configurations $F$ which would result from
successfully attempting each of the allowed moves for site $j$.
The probability to undergo each transition is then calculated from Eq.~(\ref{eq:dynrate}).
Blocked moves have transition probability zero.
To ensure microscopic reversibility (detailed balance), the possibility that no
change in configuration occurs ($F=I$) must also be included in the list of new configurations.
The probability to remain in the initial state is $R(I|I)=1-\sum_{F \ne I}R(F|I)$.

In general, there is a possibility that $R(I,I)$ could be negative.
However, this essentially means that the time corresponding to one MCSS is chosen too long for a 
consistent stochastic description.
Thus, $\nu$ must be chosen sufficiently small that $R(I|I)$ remains positive.
As long as this condition is satisfied, changing $\nu$ only 
corresponds to changing the overall time scale for the simulation.
In our simulations we used $\nu=1$~(MCSS)$^{-1}$, which 
ensured that $R(I|I)$ remained positive with the barriers used here.

The main price paid for the simplicity of coding our algorithm is that
it is rather slow in parameter regimes where the acceptance
probabilities are very low. However, 
except for extremely {\it fast} processes, 
in which the acceptance probabilities are large and  
change significantly within a single MCSS, it is equivalent to 
more sophisticated continuum-time algorithms 
\cite{KANG89,BORT75,GILM76,FICH91,NOVO95,LUKK98}. 

The adsorption/desorption barrier is $\Delta_{\rm a}$,
the nearest-neighbor diffusion barrier is $\Delta_{\rm nn}$,
and the next-nearest neighbor diffusion barrier is $\Delta_{\rm nnn}$.
In the simulations, the diffusion barriers are 
$\Delta_{\rm nn}=100$~meV and $\Delta_{\rm nnn}=200$~meV,
based on {\it ab-initio} calculations \cite{IGNACZAK:QMHALIDE} 
of binding energies for a single Br ion to a Ag(100) substrate in vacuum.
The barrier $\Delta_{\rm nn}$ is approximated by the difference 
in binding energy between the four-fold hollow site and the bridge site, 
while $\Delta_{\rm nnn}$ is approximated by the difference in binding energy between the four-fold
hollow site and the on-top site.

Theoretical estimates of $\Delta_{\rm a}$ are extremely sensitive to the
details of both the ion-surface and ion-water interactions \cite{IGNA98}. 
Based on data from calculations of the potential of mean force for halide ions in water 
near a Cu(100) surface \cite{IGNA98}, our very rough estimates indicate that
values between 200 and 500~meV are not unreasonable. 
Since we also expect $\Delta_{\rm a}$ to be significantly larger than $\Delta_{\rm nnn}$, 
we examined some aspects of the behavior for $\Delta_{\rm a}=300$~meV and $\Delta_{\rm a}=400$~meV. 
Finding few qualitative differences between simulation results for these two values,  
we chose the lower value, $\Delta_{\rm a}=300$~meV, to increase the simulation speed. 

\section{Simulated Potential Steps}
\label{sec:step}

One way to probe the dynamics of an adsorbed layer is to subject it 
to a sudden change in electrochemical potential.
Dynamic simulations of such potential-step experiments are started by equilibrating the 
room-temperature system at a potential $\bar{\mu}_1$, using the simulation technique described in Sec.~3.
Then, at time $t=0$, the potential is instantaneously stepped to a new value $\bar{\mu}_2$,
and the simulation continues by the dynamic MC algorithm described in Sec.~\ref{sec:dynMC}.
To study the dynamics of the phase ordering and disordering processes,
we examined four different potential steps: 
two from the disordered phase into the ordered phase, 
and two from the ordered phase into the disordered phase.

Before discussing each of these potential steps in detail,
we give the initial adsorption, desorption, and diffusion rates at 
$\bar{\mu}_2$, immediately after the sudden potential change.
We define the probability during one MCSS of accepting (when possible) an adsorption move as $R_{\rm a}$, 
a desorption move as $R_{\rm d}$,
a nearest-neighbor diffusion move as $R_{\rm nn}$,
and a next-nearest neighbor diffusion move as $R_{\rm nnn}$.
As an example, we present the rate calculations for a deep disorder-to-order step 
with $\bar{\mu}_1=-200$~meV and $\bar{\mu}_2=+600$~meV. 
Each of the four rates is calculated from Eq.~(\ref{eq:dynrate}).
The initial and final energies, $U_I$ and $U_F$, respectively, are calculated from Eq.~(\ref{eq:hamil}).
At $\bar{\mu}_1$, $\Theta\approx 0$; therefore, at early times the surface is nearly empty.
For adsorption, $I$ is approximated as having $\Theta=0$,
and $F$ contains only one occupied site.
Thus, $U_I=0$, $U_F=-\bar{\mu}_2$, and $R_{\rm a}=\nu \exp{[(-\Delta_{\rm a} -(U_F-U_I)/2)/k_{\rm B} T]}=\nu \exp{[0.0]}$.
For desorption, $I$ must contain at least one occupied site for the process to be possible;
therefore, $I$ is approximated as having only one occupied site,
and $F$ is approximated as an empty surface.
Thus, $U_I=-\bar{\mu}_2$, $U_F=0$, and $R_{\rm d}=\nu \exp{[(-\Delta_{\rm a} -(U_F-U_I)/2)/k_{\rm B} T]}=\nu \exp{[-24.0]}$.
For nearest-neighbor diffusion, $I$ is also approximated as having only one occupied site,
and $F$ is one of the four possible configurations which would occur from a nearest-neighbor diffusion move.
Thus, $U_I=U_F=-\bar{\mu}_2$, and 
$R_{\rm nn}=4 \nu \exp{[(-\Delta_{\rm nn} -(U_F-U_I)/2)/k_{\rm B} T]}=4 \nu \exp{[-4.0]}$,
where the factor of 4 comes from adding the contributions from each of the four final states.
Similarly, for next-nearest neighbor diffusion, $U_I=U_F=-\bar{\mu}_2$, and 
$R_{\rm nnn}=4 \nu \exp{[(-\Delta_{\rm nnn} -(U_F-U_I)/2)/k_{\rm B} T]}=4 \nu \exp{[-8.0]}$.
The initial rates for all four potential steps are summarized in Table 1.
Note that the rates change significantly during the course of the time evolution following each step.

\begin{figure}
{\epsfxsize=6in \epsfbox{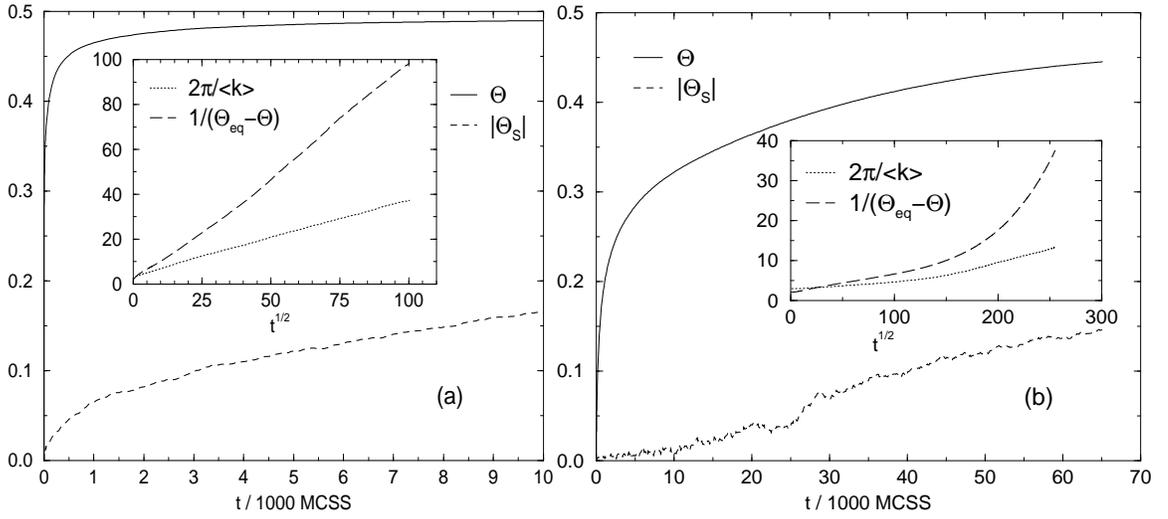}}
\caption[]{
Time series for sudden disorder-to-order potential-steps at room temperature
for $L=256$ and $\phi_{\rm nnn}=-26$~meV.
The insets show the correlation length for the c$(2\times 2)$ order-parameter 
fluctuations vs.\ $t^{1/2}$.
(a): Deep step from 
$\bar{\mu}_1=-200$~meV to $\bar{\mu}_2=+600$~meV,
averaged over 10 independent runs.
(b): Shallow step from 
$\bar{\mu}=-200$~meV to $\bar{\mu}_{\rm 2}=+250$~meV,
averaged over 7 independent runs.
}
\label{fig:posstep}
\end{figure}

We now discuss each of the four potential steps.
First we consider a disorder-to-order step with $\bar{\mu}_1=-200$~meV 
and $\bar{\mu}_2=+600$~meV. 
At $\bar{\mu}_1$, $\Theta\approx 0$,
and the final potential $\bar{\mu}_2$ is deep into the ordered phase, 
approximately $ 420$~meV on the positive side of the transition at
$\bar{\mu}_c \approx 180$~meV.
Figure~\ref{fig:posstep}(a) shows time series for both $\Theta$ and $|\Theta_{\rm S}|$.
For deep steps like these, the desorption rate is negligible (see Table~1). 
Thus, the adsorption dynamics are essentially described by the process of random sequential
adsorption with diffusion (RSAD) \cite{WANG93,EISE98}.
The coverage quickly reaches the jamming coverage for
random sequential adsorption of hard squares without diffusion, 
$\Theta_{\rm J} \approx 0.364$
\cite{MEAK87,BARA95R,EVAN93}\footnote{
Precise values for $\Theta_{\rm J}$ in the hard-square model, 0.364\,13(1)
and 0.364\,133\,0(5), have been obtained by MC simulations
\protect\cite{MEAK87} and series expansions \protect\cite{BARA95R},
respectively.}.
This is only slightly less than the critical coverage, $\Theta_c \approx 0.368$, 
which is reached in approximately 37~MCSS. 
During this time, small, uncorrelated domains of the ordered phase are formed on both sublattices.
As the lattice fills up, the interfaces between these domains contain empty sites, 
but the nearest-neighbor exclusion generally prevents the adsorption of additional Br ions.
This is followed by the growth of adsorbate domains on the two sublattices which is well
described by dynamical scaling (see below). 
At this stage, the coverage can increase only where domain walls move together, 
opening a gap large enough to fit an additional Br.
In the absence of single-particle diffusion, the domain walls can move only by
a series of very improbable desorption/adsorption events.
However, nearest-neighbor diffusion allows some Br ions to diffuse between adjacent domains,
thus providing a much faster mechanism for interface motion.
For deep steps with the barriers used here, interface motion is almost exclusively due
to nearest-neighbor diffusion. 
Excluding very small fluctuations on short time scales, the coverage is always driven 
toward the equilibrium value, $\Theta_{\rm eq}=1/2$.
Thus, the total length of interfaces is steadily decreasing.
Finally, selection of one or the other of the two degenerate sublattices takes place at late times.
The phase-selection time is expected to depend on $L$.

\begin{figure}
{\epsfxsize=6in \epsfbox{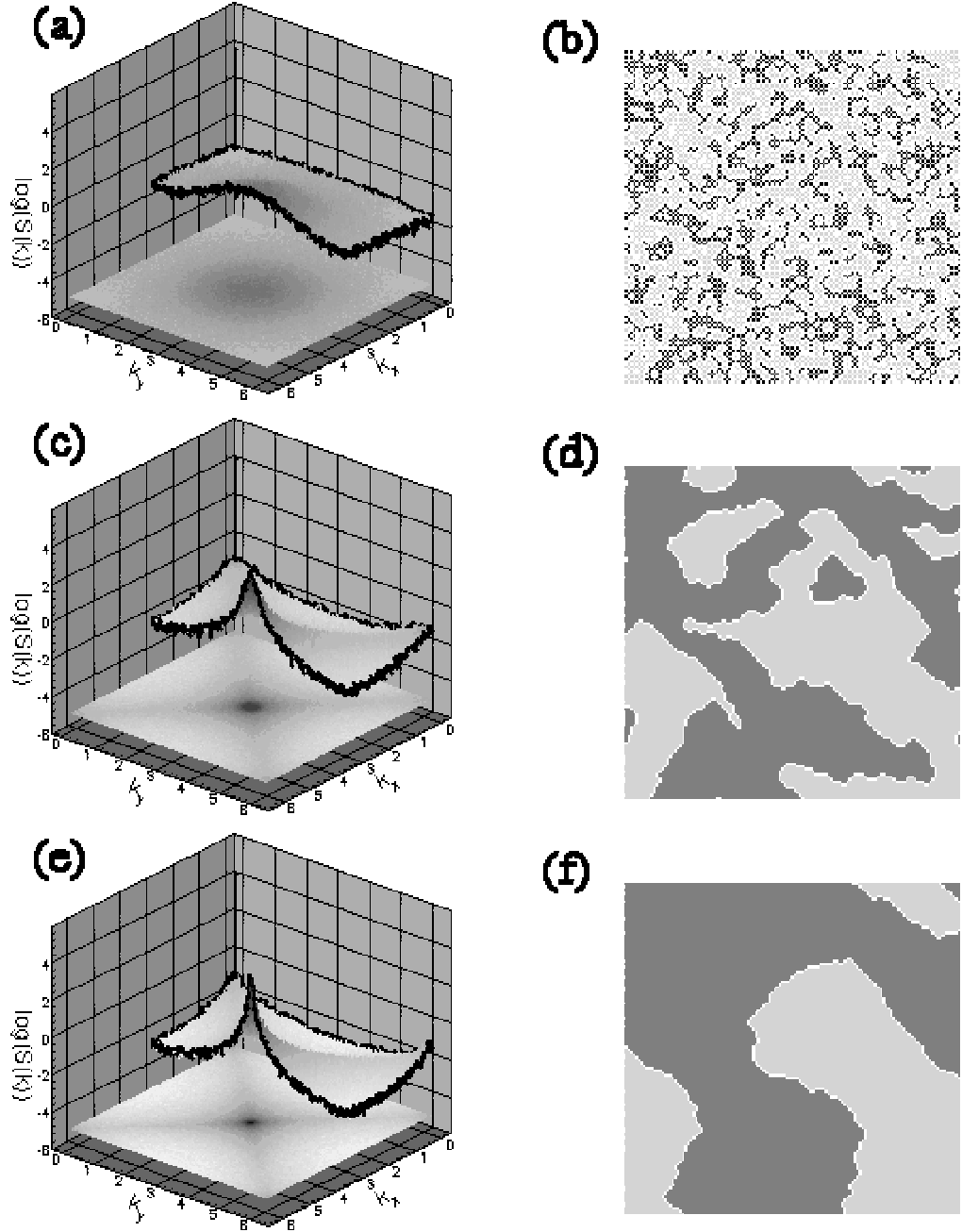}}
\caption[]{
Sequence of structure factors and real-space configurations
following a deep disorder-to-order step, $\bar{\mu}_1=-200$~meV, $\bar{\mu}_2=+600$~meV,
$L$=256, $T\approx$290~K.
(a,c,e): Structure factor in three-dimensional cut-away and full projected views
at times 25, 2000, and 10,000~MCSS, respectively, averaged over 10 independent runs.
The grayscale on each is the same.
(b,d,f): 128$\times$128 section of a 256$\times$256 
real-space configuration at times
25, 2000, and 10,000~MCSS, respectively.
The two different shades of gray idicate Br on the two sublattices.
}
\label{fig:S3D}
\end{figure}

\begin{figure}
{\epsfxsize=6in \epsfbox{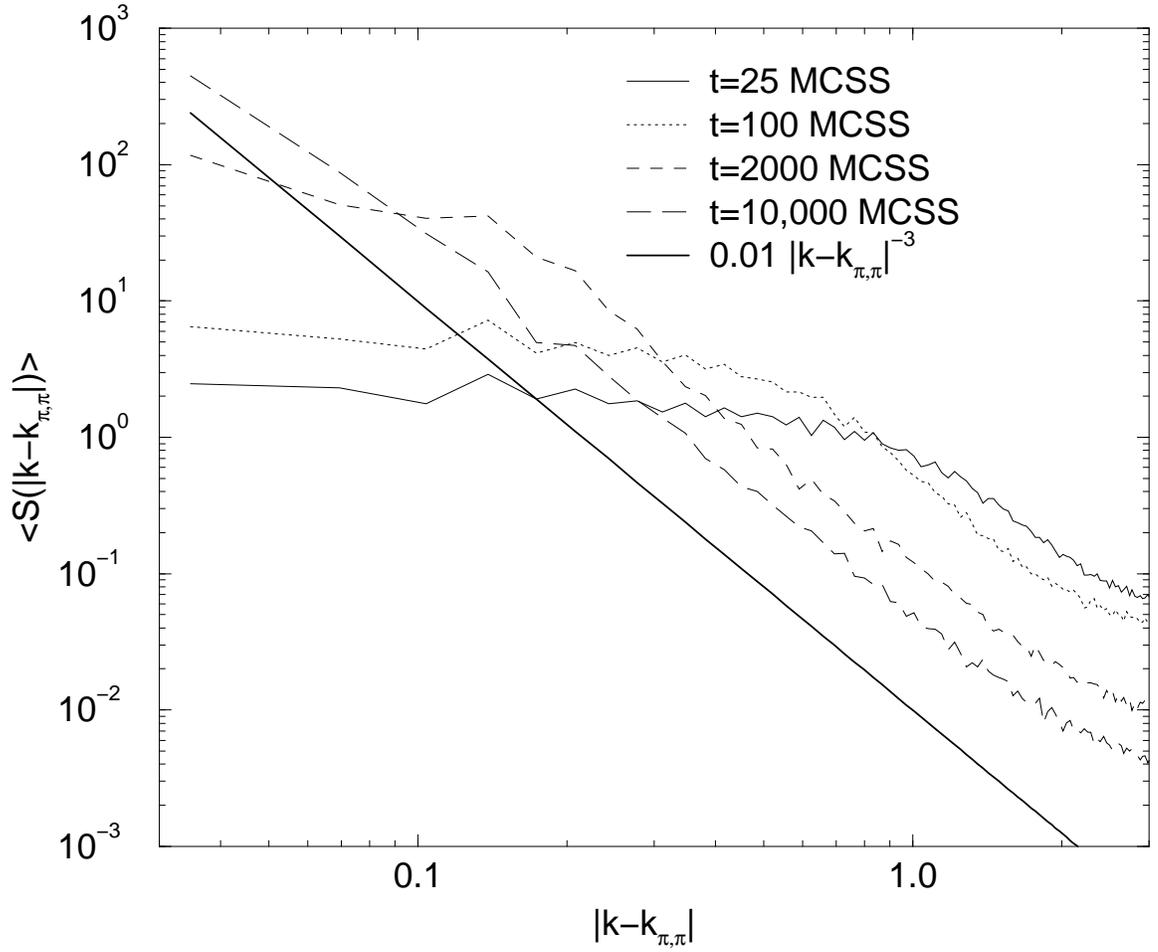}}
\caption[]{
Circularly averaged structure 
factor at various times for a deep disorder-to-order
step from $\bar{\mu}_1=-200$~meV
to $\bar{\mu}_2=+600$~meV for $L=256$ and $\phi_{\rm nnn}=-26$~meV.
For comparison with Porod's Law \cite{POROD,GUINIER}, the heavy line indicates the power law $|\vec{k}-\vec{k}_{\pi,\pi}|^{-3}$.
}
\label{fig:Save}
\end{figure}

In the dynamical scaling regime, the correlation length $\xi$ of the system 
should grow as a power law with time $t$ \cite{GUNTON:DYNTRAN},
\begin{equation}
\xi\propto t^n.
\end{equation}
In models with non-conserved order parameter such as the one studied here, 
the domain growth is driven solely by the tendency to reduce the amount of interface.
For such systems (known as Model A in the nomenclature of critical dynamics \cite{GUNTON:DYNTRAN,HOHE77})
$n=1/2$ \cite{GUNTON:DYNTRAN,HOHE77,LIFS62,ALLE79}.
We have used two different measures for $\xi$.
One common measure uses the width of the diffuse structure-factor peak corresponding to the ordered phase,
here the $(1/2,1/2)$ or $\vec{k}=(\pi,\pi)$ peak.
We define the correlation length given by the 
structure factor as $\xi_S=2\pi/\langle k\rangle$,
where 
\begin{equation}
\langle k\rangle=
\frac{\sum_{|\vec{k}-\vec{k}_{\pi,\pi}|>0}^{\prime} 
|\vec{k}-\vec{k}_{\pi,\pi}| \langle S(|\vec{k}|)\rangle}
{\sum_{|\vec{k}-\vec{k}_{\pi,\pi}|>0}^{\prime} \langle S(|\vec{k}|)\rangle} \;.
\end{equation}
Here $\langle S(|\vec{k}|)\rangle$ is the circularly averaged $S(\vec{k})$
and $\sum_{|\vec{k}-\vec{k}_{\pi,\pi}|>0}^{\prime}$ runs over all allowed
values of $|\vec{k}| \ne |\vec{k}_{\pi,\pi}|$, 
such that $|\vec{k}-\vec{k}_{\pi,\pi}|$ is less than $\pi$
to avoid picking up contributions from the small diffuse scattering 
surrounding the $\vec{k}=(0,0)$ Bragg peak. 
Figure~\ref{fig:S3D} shows three-dimensional 
views of the structure factor and typical system configurations
at three different times following this deep step
into the c$(2 \times 2)$ ordered phase.
The diagonal anisotropy in $S(\vec{k})$ at late times
corresponds to domain walls oriented at 45$^{\circ}$ to the lattice directions,
which are stable under diffusion.
Figure~\ref{fig:Save} shows the circularly averaged structure factor at various times.
At late times the structure factor is proportional to $|\vec{k}-\vec{k}_{\pi,\pi}|^{-3}$.
This asymptotic behavior, known as Porod's law \cite{POROD,GUINIER}, indicates that the domain
walls are microscopically thin.

For this system, we are also 
able to exploit the area covered by the empty domain walls as a second 
measure of $\xi$ \cite{WANG93,EISE98},
since a decrease in interfacial length is equivalent to an increase 
in the correlation length \cite{DEBY57}.
We define this second measure as $\xi_W=(\Theta_{\rm eq}-\Theta)^{-1}$, where 
$\Theta_{\rm eq}$ is the equilibrium value of the coverage at $\bar{\mu}_2$ 
(here, $\Theta_{\rm eq} = 1/2$). The inset in Fig.~\ref{fig:posstep}(a) 
shows $\xi_S$ and $\xi_W$ as functions of $t^{1/2}$ for $L=256$. 
The agreement with the expected $t^{1/2}$ behavior is excellent. 
This behavior has also been found in previous studies of RSAD \cite{WANG93,EISE98}.
For an infinite system, $\xi$ would continue to grow without bound;
however, when $\xi$ reaches the same order of magnitude as $L$,
the system enters the phase-selection regime in which one of the
two degenerate ordered phases is randomly selected.

For the other disorder-to-order step simulation, we take $\bar{\mu}_1=-200$~meV and 
$\bar{\mu}_2=+250$~meV. This step ends approximately 
$70$~meV on the positive side of the transition.
Figure~\ref{fig:posstep}(b) shows the dynamic 
behavior following this shallow step into the ordered phase.
As one would expect, the overall response  
is much slower than for the deep step. 
Although the behavior at early times is more complicated than for the deep step, 
$\xi_S$ does show $t^{1/2}$ domain growth at late times (see inset). 
The length scale evaluated from the coverage, $\xi_W$, is not an accurate measure 
for shallow steps in the time range we have simulated,
since the interfaces are not as clearly defined as in the deep step.

As expected, the ordering dynamics are well described
by the dynamical scaling prediction of Model A.
Although dynamical scaling has yet to be experimentally measured in this system,
it has been measured in other electrochemical adsorption systems \cite{FINN98},
and we believe that $t^{1/2}$ dynamical scaling will
be experimentally observed in the future for this system as well.

\begin{figure}
{\epsfxsize=6in \epsfbox{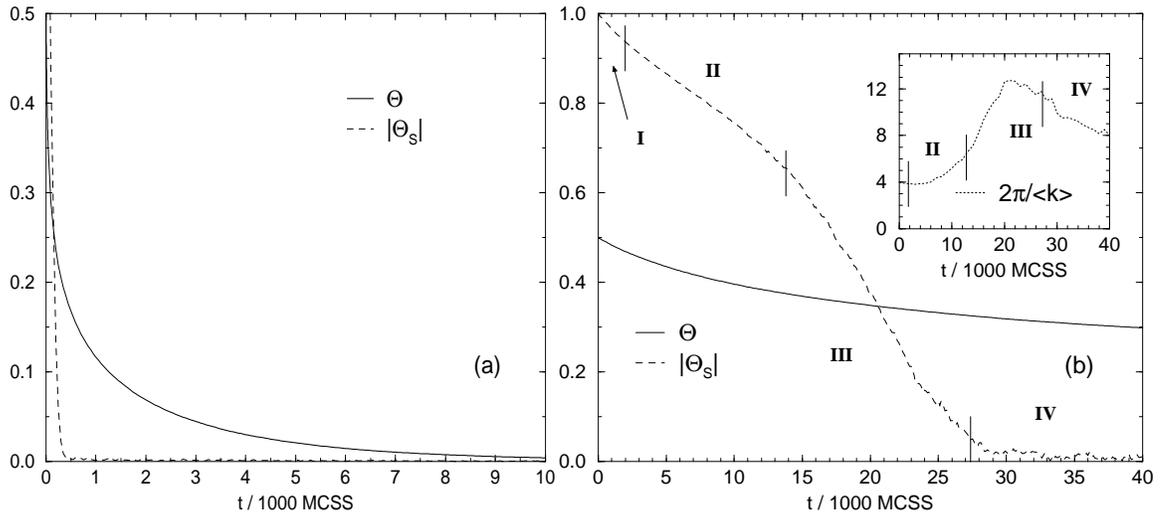}}
\caption[]{
Time series for sudden order-to-disorder potential steps at room temperature
for $L=256$ and $\phi_{\rm nnn}=-26$~meV.
(a): Deep step from 
$\bar{\mu}_1=+600$~meV to $\bar{\mu}_2=-200$~meV,
averaged over 3 independent runs. 
(b): Shallow step from $\bar{\mu}_1=+600$~meV 
to $\bar{\mu}_2=+100$~meV, averaged over 4 independent runs.
The inset shows the correlation length $\xi_S$ 
for the c$(2\times 2)$ order-parameter fluctuations vs.\ $t$.
All four dynamical regimes discussed in the text are labeled with Roman numerals.
}
\label{fig:negstep}
\end{figure}

Potential steps that induce disordering are also of interest. 
For deep negative steps, $\bar{\mu}_1$=$+600$~meV and $\bar{\mu}_2$=$-200$~meV,
the disordering process is relatively uneventful, 
as shown in Fig.~\ref{fig:negstep}(a). 
Particles simply desorb at a roughly constant rate,
leading to an essentially exponential relaxation to equilibrium.

Much more interesting behavior is seen after a shallow negative step, 
such as $\bar{\mu}_1$=$+600$~meV and $\bar{\mu}_2$=$+100$~meV, shown in Fig.~\ref{fig:negstep}(b).
The simulation begins with all particles on sublattice $A$ and relaxes to the
disordered phase at $\bar{\mu}_2$, where $\Theta\approx 1/4$.
Four different dynamical regimes are seen.
In the first regime, particles simply desorb from the surface,
so that ${\rm d}\Theta_{\rm S}/{\rm d}t\approx2\,{\rm d}\Theta/{\rm d}t$.
In the second regime, both desorption and diffusion contribute significantly
to the disordering process.
As more sites become vacant by desorption, 
small domains of the $B$ phase are created,
both by lateral diffusion from sublattice $A$ and by re-adsorption from the solution. 
As a result, the diffuse scattering peak narrows.
In the third regime, the net desorption slows down, 
and diffusion and to some extent adsorption/desorption are the only significant 
contributions to the disordering process.
Here, ${\rm d}\Theta_{\rm S}/{\rm d}t$ is largest.
In the final regime, the two sublattices become approximately equally populated,
yielding $\Theta_{\rm S}\approx0$.
Similar behavior is seen in a dynamic mean-field theory \cite{RIKVOLD:DMFT}.

The inset in Fig.~\ref{fig:negstep}(b) shows the correlation length obtained from  
the diffuse (1/2,1/2) scattering peak as a function of time for the step to +100~meV.
This length can be interpreted as the typical distance over which  
the order-parameter fluctuations are correlated.
Thus, at early times, when the Br are predominantly on sublattice $A$,
the fluctuations onto sublattice $B$ are very small. 
We believe that the temporary increase in the correlation length 
obtained from the diffuse scattering
is caused by the system passing near the critical point of the phase transition. 
The length scale of the fluctuations decreases again as the 
coverage falls below its critical value and the macroscopic order is lost.

\begin{figure}
{\epsfxsize=6in \epsfbox{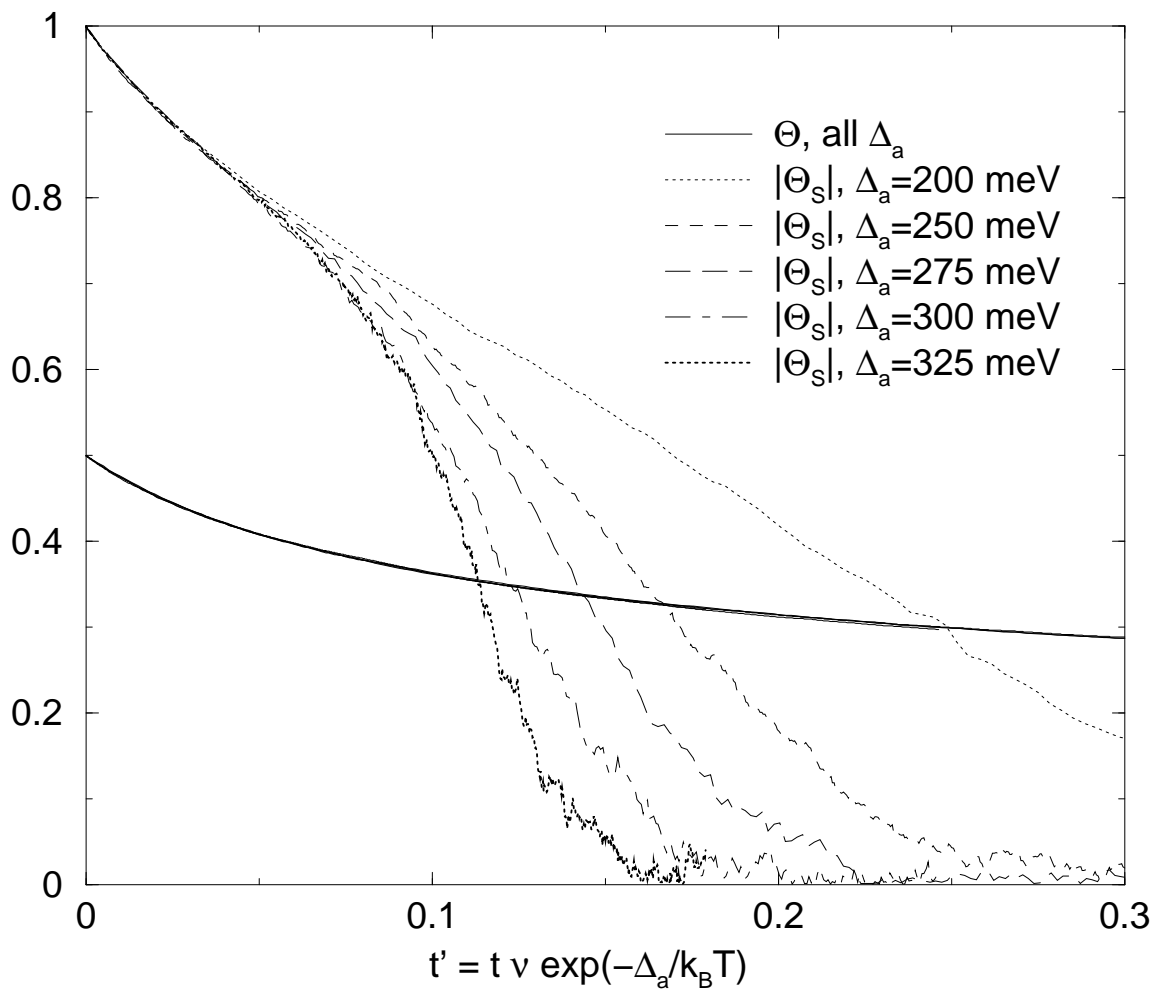}}
\caption[]{
Time series for shallow negative potential step with various values of the adsorption/desorption
barrier $\Delta_{\rm a}$.
All other parameters have the same values as in Fig.~\ref{fig:negstep}.
Time is given in dimensionless units as $t'=t \nu \exp{(-\Delta_{\rm a}/k_B T)}$.
The data show the results of a single potential-step simulation for each value of $\Delta_{\rm a}$.
}
\label{fig:rsqdiffbar}
\end{figure}

\begin{figure}
{\epsfxsize=6in \epsfbox{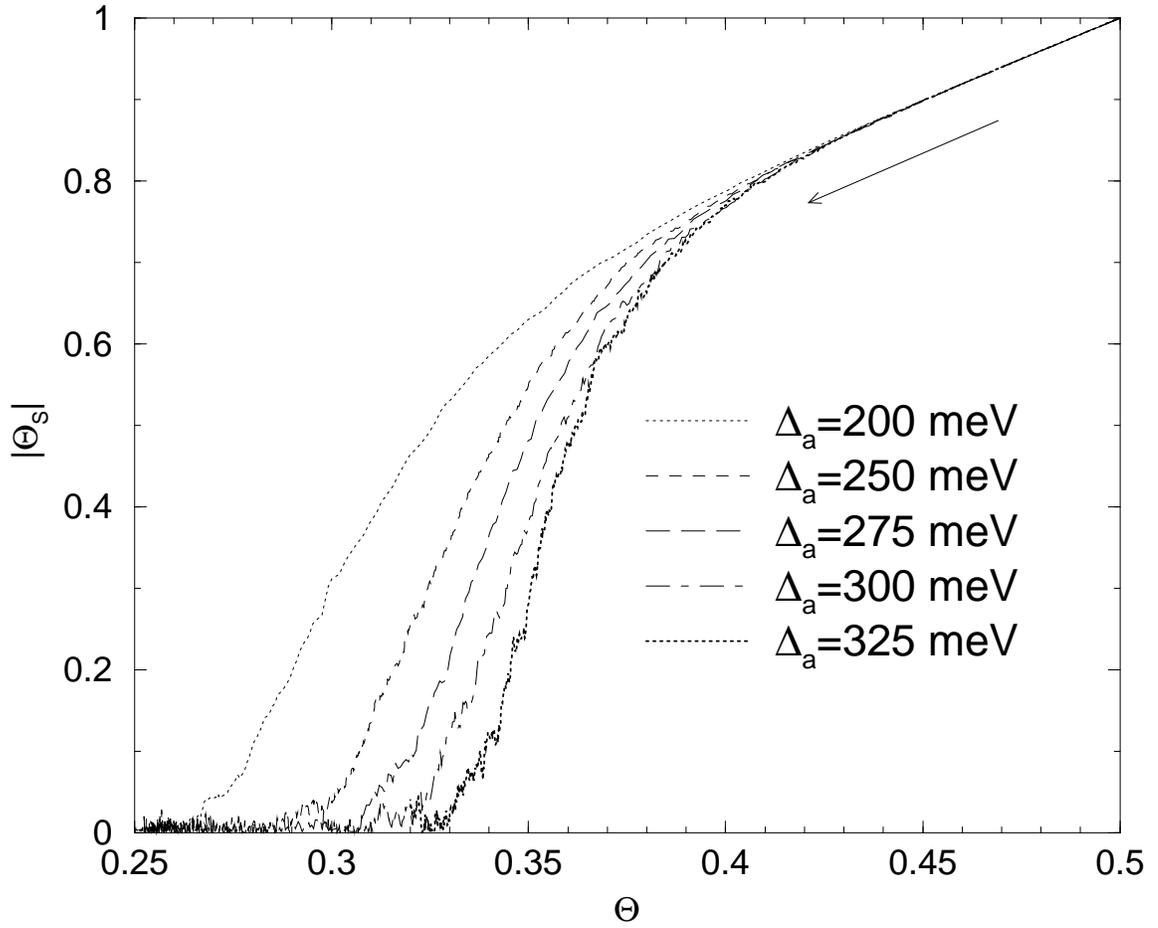}}
\caption[]{
Parametric plot of $|\Theta_{\rm S}|$ vs.\ $\Theta$ for a shallow negative potential-step 
with various values of the adsorption/desorption barrier $\Delta_{\rm a}$.
The data shown are the same as in Fig.~\ref{fig:rsqdiffbar}.
The arrow indicates the direction of increasing time.
}
\label{fig:parametric}
\end{figure}

For shallow disordering steps, time series for $\Theta$ and $\Theta_{\rm S}$ 
are quite sensitive to the choice of the adsorption/desorption barrier.
Here we discuss how this dependence can be exploited
to estimate both $\Delta_{\rm a}$ and the rate constant $\nu$ from future experimental results.
For simplicity, we assume that our choices of diffusion barriers are realistic.
Neglecting this assumption does not affect our qualitative conclusions
since the time scale for the disordering process is dominated by $\Delta_{\rm a}-\Delta_{\rm nn}$.
We first discuss $\Theta(t')$, which depends on time
only through the dimensionless rescaled time $t'=t \nu \exp{(-\Delta_{\rm a}/k_{\rm B} T)}$.
See Fig.~\ref{fig:rsqdiffbar}.
By comparing experimental measurements of $\Theta(t)$ with the rescaled simulation results,
the product $\nu \exp{(-\Delta_{\rm a}/k_{\rm B} T)}$ can be estimated.
However, knowledge of $\Theta(t)$ alone is not enough to independently
estimate both $\nu$ and $\Delta_{\rm a}$.
To estimate both parameters, knowledge of $|\Theta_{\rm S}(t)|$ is also required.
The dependence of $|\Theta_{\rm S}|$ on $\Delta_{\rm a}$ can be seen in Fig.~\ref{fig:rsqdiffbar},
but the dependence is more readily seen in a parametric plot of $|\Theta_{\rm S}(\Theta)|$, which is independent of $\nu$.
Such a plot is shown in Fig.~\ref{fig:parametric}.
In this form, the shape dependence on $\Delta_{\rm a}$ is more obvious.
Also, the result that ${\rm d}|\Theta_{\rm S}|/{\rm d}t=2{\rm d}\Theta/{\rm d}t$
at short times is easily seen in the parametric plot.
By plotting $|\Theta_{\rm S}(\Theta)|$ from experiments and simulations together,
it should be possible to estimate $\Delta_{\rm a}$, 
assuming that the diffusion barriers used are realistic.
Once $\Delta_{\rm a}$ has been estimated, the product $\nu \exp{(-\Delta_{\rm a}/k_{\rm B} T)}$,
known from comparing the time scales of the experimental and simulated processes,
can be used to estimate $\nu$.
It is our hope that experimental measurement of both $\Theta(t)$ and $|\Theta_{\rm S}(t)|$
will be possible in the future for this shallow disordering process.

\section{Simulated Cyclic Voltammograms}
\label{sec:CV}

\begin{figure}
{\epsfxsize=6in \epsfbox{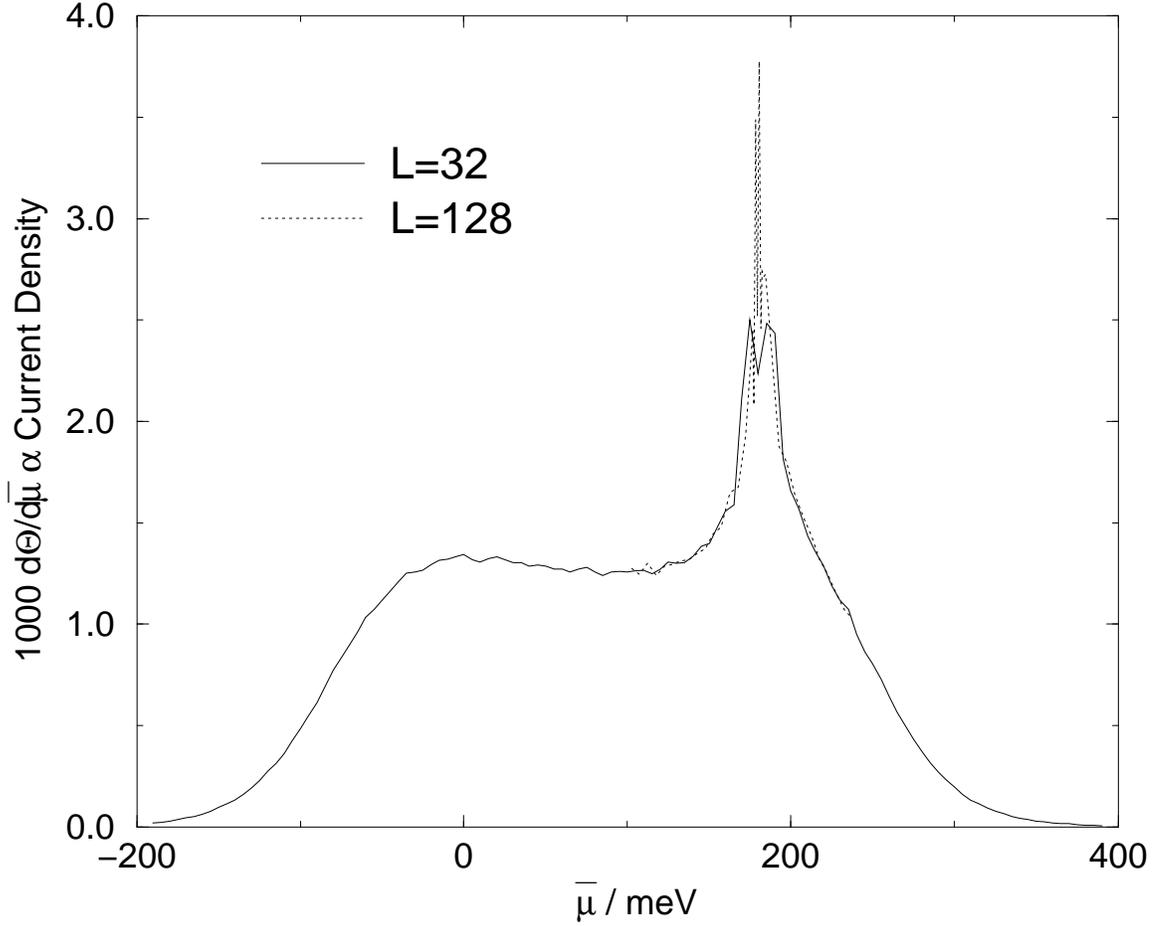}}
\caption[]{
Quasi-equilibrium CV for $L=32$ and $L=128$ at $T\approx 290$~K with $\phi_{\rm nnn}=-26$~meV.
The derivative ${\rm d}\Theta/{\rm d}\bar{\mu}$ is calculated using a three-point difference method.
The peak corresponds to the second-order phase transition 
between the disordered and c$(2 \times 2)$ phases.
The broad shoulder on the negative side of the sharp peak
is due to configurational fluctuations of the 
Br adlayer as discussed in the text.
The statistical errors are larger for $L=128$ than for $L=32$.
}
\label{fig:qCV}
\end{figure}

\begin{figure}
{\epsfxsize=6in \epsfbox{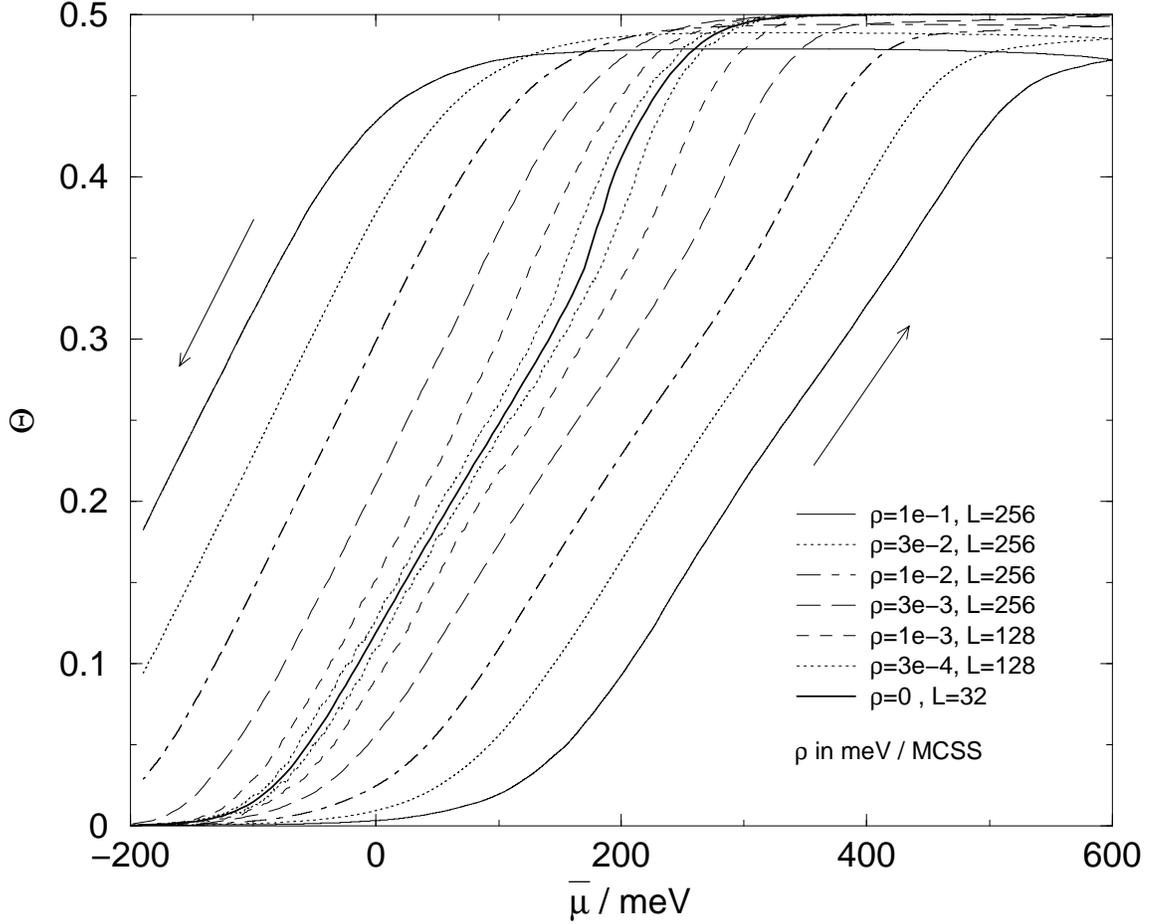}}
\caption[]{
$\Theta$ vs.\ $\bar{\mu}$ for various sweep rates $\rho$.
Dynamic MC simulations were used to generate the curves for $\rho\ne0$.
For $\rho$=0, the equilibrium coverage from Fig.~\ref{fig:isoth} is shown.
$L$=256 for $\rho$=$1$$\times$$10^{-1}$ meV/MCSS to $3$$\times$$10^{-3}$ meV/MCSS,
$L$=128 for $\rho$=$1$$\times$$10^{-3}$ and $3$$\times$$10^{-4}$ meV/MCSS, and
$L$=32 for $\rho$=0.
Here, $\bar{\mu}_1$=$-200$~meV for $\rho$=$1$$\times$$10^{-1}$ meV/MCSS to $1$$\times$$10^{-2}$ meV/MCSS
and $\bar{\mu}_1$=$-400$~meV otherwise, while $\bar{\mu}_2=+600$~meV.
The arrows indicate the direction of increasing time.
}
\label{fig:CVcov}
\end{figure}

\begin{figure}
{\epsfxsize=6in \epsfbox{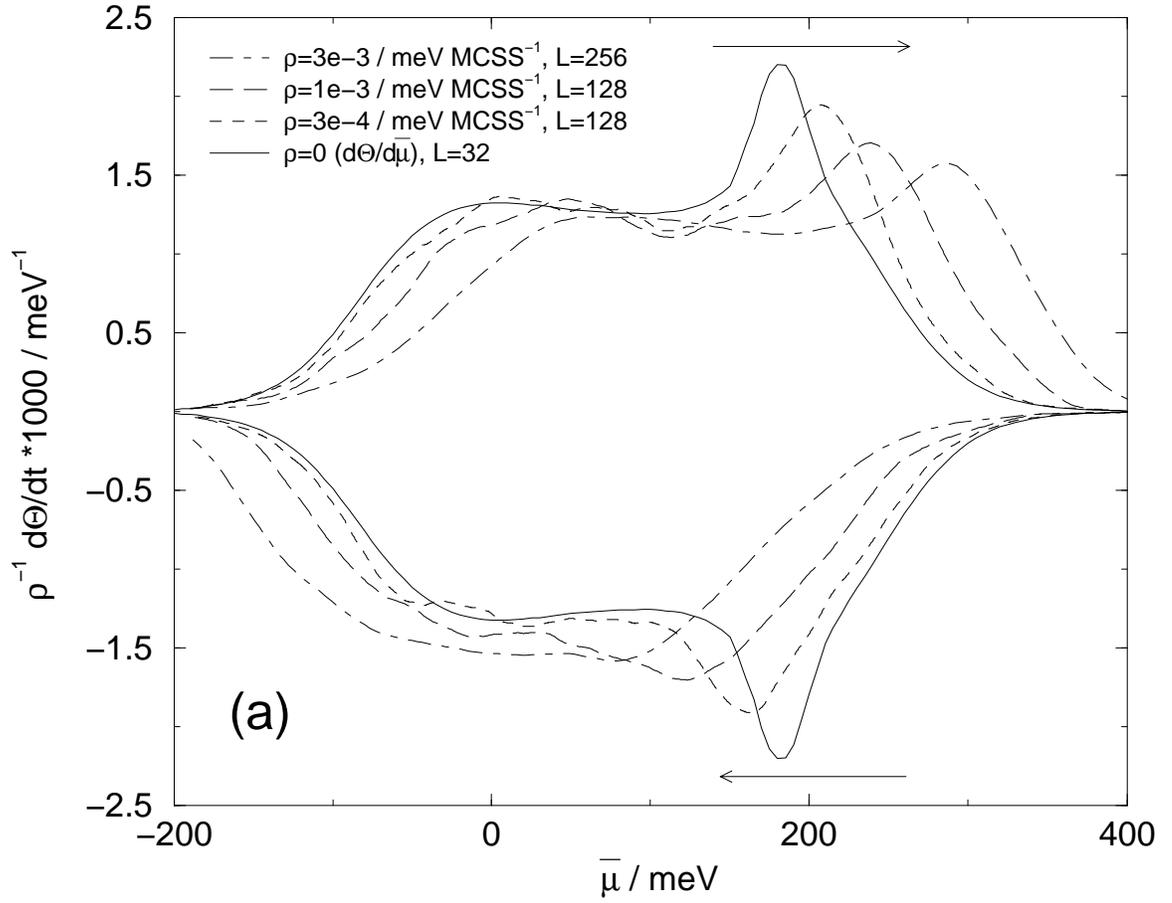}}
\caption[]{
Dynamically simulated cyclic voltammograms (CVs) for various sweep rates 
showing ${\rm d}\Theta/{\rm d}t$ vs.\ $\bar{\mu}$.
For all non-zero sweep rates, the dynamic response ${\rm d}\Theta/{\rm d}t$,
which is proportional to the voltametric current,
is normalized by the sweep rate.
For $\rho$=0, the quasi-equilibrium CV, ${\rm d}\Theta/{\rm d}\bar{\mu}$ from Fig.~\ref{fig:qCV}, is shown.
All parameters and symbols are as in Fig.~\ref{fig:CVcov}.
}
\label{fig:CV}
\end{figure}

Next we use the dynamic Monte Carlo method described in Sec.~\ref{sec:dynMC}
to study the dynamic response of the adlayer to a continuously varying potential,
as realized in a CV experiment.
The system displays a phase lag between the ordering and disordering processes and the time dependent $\bar{\mu}$.
For these simulated CVs, $\bar{\mu}$ is linearly ramped 
from $\bar{\mu}_1$ to $\bar{\mu}_2$ with a sweep rate $\rho$
and then ramped back to $\bar{\mu}_1$ at the same rate.
A wide range of sweep rates were simulated,
from 10$^{-1}$ to $3 \times 10^{-4}$~meV/MCSS.
Figure~\ref{fig:CVcov} shows $\Theta$ as a function of $\bar{\mu}$
and is analogous to the familiar hysteresis plots for magnetic systems,
showing the magnetization as a function of the applied magnetic field \cite{ME:UGA00}.
The net flux of Br$^-$ into the adlayer is proportional to the voltammetric current.
Figure~\ref{fig:CV} shows $\rho^{-1} {\rm d}\Theta/{\rm d}t$ as a function of $\bar{\mu}$
for the three slowest sweep rates from Fig.~\ref{fig:CVcov},
along with the quasi-equilibrium CV from Fig.~\ref{fig:qCV}.
For the faster sweep rates, we used $L=256$,
while for the slowest sweep rates, $L=128$ was used so that simulations could be performed
in a reasonable amount of time 
(about 30 days of CPU time per cycle for $\rho=3 \times 10^{-4}$~meV/MCSS on a Pentium III 500 MHz processor).
To reduce the effects of statistical noise,
the derivatives were calculated by the Savitzky-Golay smoothing method using a quadratic fit 
to the $\Theta(t)$ data over a $\bar{\mu}$ range of about 60~meV \cite{PRES92}.
This accounts for the reduced peak height for $\rho=0$, compared to Fig.~\ref{fig:qCV}.

\begin{figure}
{\epsfxsize=6in \epsfbox{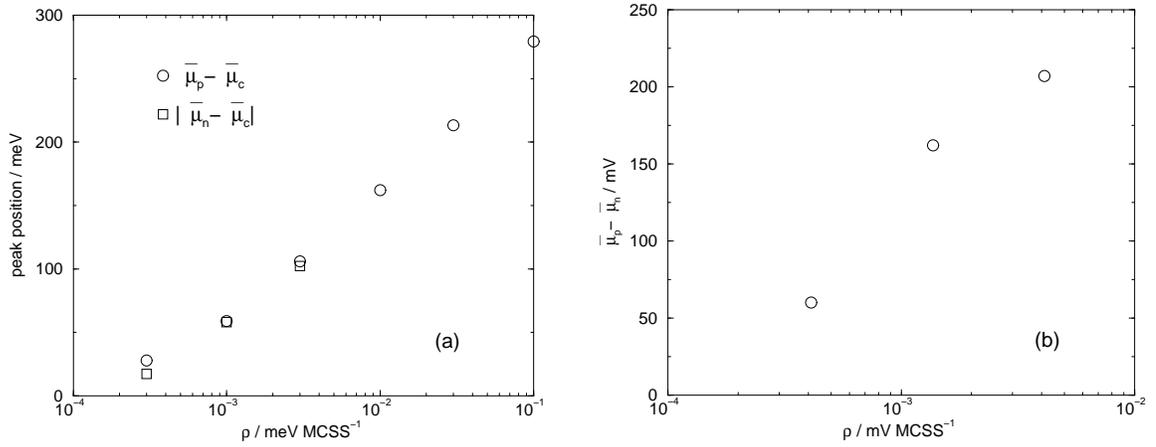}}
\caption[]{
Peak positions and peak separations for various sweep rates.
(a): Positive-going and negative-going peak positions vs.\ $\rho$.
All parameters are the same as in Figs.~\ref{fig:CVcov} and \ref{fig:CV}.
(b): Peak separations $\bar{\mu}_{\rm p}-\bar{\mu}_{\rm n}$.
For comparison with experiments, the units are given in mV using $\gamma=-0.73$.
}
\label{fig:approach}
\end{figure}

As expected, the CVs approach the quasi-equilibrium result as $\rho$ approaches zero, and
the phase lag between the response and the potential depends on the sweep rate.
For non-zero sweep rates, 
there is asymmetry between the positive-going (ordering) and negative-going (disordering) scans
caused by the differences between the phase-ordering and disordering processes
discussed in Sec.~\ref{sec:step}.
We denote the position of the sharp peak in ${\rm d}\Theta/{\rm d}t$ 
for the positive-going scan as $\bar{\mu}_{\rm p}$
and for the negative-going scan as $\bar{\mu}_{\rm n}$.
Figure.~\ref{fig:approach}(a) shows $\bar{\mu}_{\rm p}-\bar{\mu}_c$ and $|\bar{\mu}_{\rm n}-\bar{\mu}_c|$ 
as functions of $\rho$.
The peak can be distinguished in the positive-going scans for all of the simulated 
sweep rates but is only visible in negative-going scans for the three slowest rates.
We therefore report the separation between the positive-going and negative-going 
peak positions, $\bar{\mu}_{\rm p}-\bar{\mu}_{\rm n}$, for only these slowest rates, 
shown in Fig.~\ref{fig:approach}(b).
No simple behavior is seen, 
but by examining experimental peak separations \cite{OCKO:BR/AG,ENDO99,VALETTE:AG},
it is obvious that our slowest sweep rate is much faster than typical
experimental sweep rates in the range of 1-10~mV/s.
At the slowest simulated sweep rate, our peak separation is about 60~mV
(in experimental electrode-potential units).
From the shape of the curve it is tempting to suggest
that by decreasing the sweep rate by another order of magnitude,
we might obtain peak separations in the experimental range of about 10~mV.

\section{Summary and Conclusion}
\label{sec:D}

We have studied the static and dynamic behavior of Br electrosorption
on the Ag(100) single-crystal surface by Monte Carlo simulation.
We have made comparison with equilibrium experiments to determine the strength of the
lateral dipole-dipole repulsion and the electrosorption valency.
The simulations reproduce the experimentally observed second-order phase 
transition between a disordered adlayer and a doubly degenerate c$(2 \times 2)$ 
ordered phase at room temperature.
As was also found previously by Koper \cite{KOPER:HALIDE},
the simulations reproduce the broad shoulder seen in experimental cyclic voltammograms.
We explain the shoulder by small configurational fluctuations in the disordered phase
that locally resemble phases with $\Theta=1/4$, one of which is 
present for this model in equilibrium at very low temperatures.
Stronger effects of this low-temperature equilibrium phase 
might be observed in the behavior of other halides on Ag(100).

We also studied the dynamics of phase ordering and disordering processes
by stepping the potential between the disordered and ordered phases.
The ordering process is well described by $t^{1/2}$ dynamical scaling, 
consistent with growth of domains of the two degenerate c($2\times 2$) ordered phases.
For disordering, no such competition between degenerate domains exists, 
and the process is well described by random desorption with diffusion.
Study of such disordering processes may provide a way to estimate the adsorption/desorption
barrier and obtain a relation between MC time and physical time.

In simulations of CVs for a wide range of potential sweep rates
we found a pronounced asymmetry between the positive-going and negative-going scans.  
This asymmetry is caused by the differences between the ordering and disordering kinetics.
No simple behavior is seen for the peak separation as a function of the sweep rate,
but comparison with experiments suggests that our slowest sweep rate
is considerably faster than experimental sweep rates which are in the range of 1-10~mV/s.

We leave comparison of simulated and experimental dynamics for future work 
when it hopefully becomes possible to experimentally study the dynamics of this system in full detail.

\section*{Acknowledgments}

We thank J.~X.\ Wang, Th. Wandlowski, and B.~M. Ocko
for providing access to their experimental data
and for useful discussions and correspondence.
We also thank O.~M.\ Magnussen for suggesting the importance of studying 
order-to-disorder potential steps and V.~Privman for
correspondence on random sequential adsorption.
Comments on the manuscript by M.~A. Novotny and Th. Wandlowski
are also gratefully acknowledged.
Supported in part by NSF grants No.\ DMR-9634873 and DMR-9981815,
by Florida State University through
the Center for Materials Research and Technology,
the Supercomputer Computations Research Institute
(U.S.\ Department of Energy contract No. DE-FC05-85FR2500),
and the School of Computational Science and Information Technology.

\appendix
\section*{Appendix}

In this Appendix we discuss the nonlinear fitting of simulated
adsorption isotherms to the experimental data from Ref. \cite{WANG:FRUM}.

Fixing a value of $\phi_{\rm nnn}$, 
we generated a room-temperature isotherm by MC simulation,
in this Appendix denoted by $\Theta^{\rm sim}(\bar{\mu};\phi_{\rm nnn})$.
We compared this isotherm to three experimental coverage isotherms
[$\Theta^{\rm exp}(E;C_1)$, $\Theta^{\rm exp}(E;C_2)$, and $\Theta^{\rm exp}(E;C_3)$]
at three different Br$^-$ concentrations [$C_1$, $C_2$, and $C_3$].
For convenience, we set the reference concentration $C_0=1$~mM,
corresponding to the units used for the experimental KBr concentrations.
Equation (2) was inverted to obtain $E(\bar{\mu};\bar{\mu}_0,\gamma;C)$,
assuming trial values of the parameters $\bar{\mu}_0$ and $\gamma$,
and the $\bar{\mu}$-axis was rescaled into an $E$-axis for each concentration.
Thus, we obtained three simulated approximations to the individual experimental isotherms
[$\Theta^{\rm sim}(E;\phi_{\rm nnn},\gamma,\bar{\mu}_0;C_1)$,
$\Theta^{\rm sim}(E;\phi_{\rm nnn},\gamma,\bar{\mu}_0;C_2)$, and
$\Theta^{\rm sim}(E;\phi_{\rm nnn},\gamma,\bar{\mu}_0;C_3)$].

In order to obtain the ``best'' values of the parameters $\phi_{\rm nnn}$,
$\gamma$, and $\bar{\mu}_0$, we defined the simultaneous 
squared difference between the simulation at $\phi_{\rm nnn}$ 
and the experimental isotherms,
\begin{equation}
\chi^2 (\phi_{\rm nnn},\gamma,\bar{\mu}_0)
=
\sum_{k=1}^3\sum_{l=1}^{l_{\rm max}(k)} 
[\Theta^{\rm sim}(E_l^k;\bar{\mu}_0,\gamma,\phi_{\rm nnn};C_k)
- \Theta^{\rm exp}(E_l^k;C_k)]^2\;.
\end{equation}
Here $C_k$ is the $k^{th}$ experimental concentration value,
$E_l^k$ is the $l^{th}$ value of the electrode potential for the $k^{th}$ experimental isotherm, 
$\sum_l$ is the sum over all experimental data points in an isotherm,
and $\sum_k$ is the sum over the three experimental concentrations.
The $\chi^2$ calculation requires that the simulation and experimental 
isotherms be evaluated at the same $E$ values.
This was accomplished by linear interpolation of the simulation data.

\begin{figure}
{\epsfxsize=6in \epsfbox{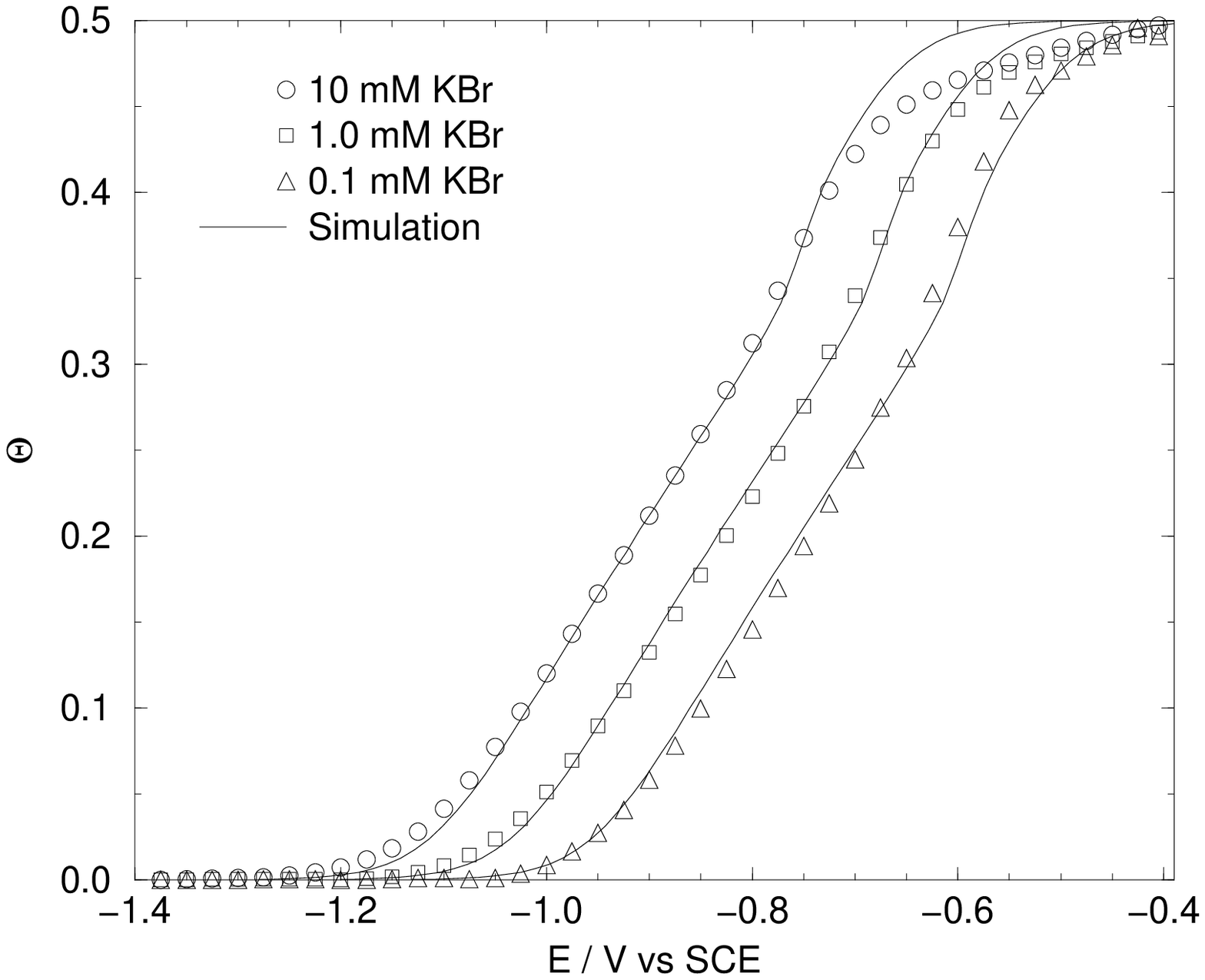}}
\caption[]{
A single Monte Carlo (MC) 
simulation isotherm ($L=32$, $T\approx 290$~K) fit simultaneously to 
chronocoulometric coverage 
isotherms at three different Br$^-$ concentrations \cite{OCKO:BR/AG}.
The parameters, $\phi_{\rm nnn}=-26\pm2$~meV, $\gamma=-0.73\pm0.03$,
and $\bar{\mu}_0=0.67\pm0.03$~meV, 
were obtained by a non-linear fitting procedure
described in the Appendix.
The reference concentration is set to $C_0=1$~mM,
and the electrode potential is given relative to the Saturated Calomel
Electrode (SCE).
}
\label{fig:exp}
\end{figure}

The parameters $\phi_{\rm nnn}$, $\gamma$, and $\bar{\mu}_0$ can, in principle, be found by minimizing 
$\chi^2$ with respect to all three parameters.
However, generating a simulated isotherm for a particular value of $\phi_{\rm nnn}$
is too computationally intensive to allow simultaneous minimization of $\chi^2$ 
with respect to all three parameters.
We therefore minimized $\chi^2$ in two steps, which saves time and allows better control
over the fitting process.
First, we generated a family of simulation isotherms 
for $\phi_{\rm nnn}$ between $-$98 and $-$10~meV at intervals of 2~meV with a 
$\bar{\mu}$ resolution of 50~meV over the electrochemical potential range $-$200 to $+1000$~meV.
For each of these simulated isotherms, 
we minimized $\chi^2$ with respect to $\gamma$ and $\bar{\mu}_0$ 
and denoted the minimum 
error obtained by 
$\chi_{\rm min}^2(\phi_{\rm nnn},\gamma^{\rm min},\bar{\mu}_0^{\rm min})$,
where $\gamma^{\rm min}(\phi_{\rm nnn})$ and 
$\bar{\mu}_0^{\rm min}(\phi_{\rm nnn})$ are the values of $\gamma$ and 
$\bar{\mu}_0$ 
which minimize $\chi^2$ for a particular value of $\phi_{\rm nnn}$.
In the second step, we minimized 
$\chi_{\rm min}^2(\phi_{\rm nnn},\gamma^{\rm min},\bar{\mu}_0^{\rm min})$
with respect to $\phi_{\rm nnn}$.
The final parameter results are $\phi_{\rm nnn}=-26\pm2$~meV,
$\gamma=-0.73\pm0.03$, and $\bar{\mu}_0=0.67\pm0.03$.
These errors are estimated from the $\phi_{\rm nnn}$
resolution and the differences in best-fit parameter
values around $\phi_{\rm nnn}=-26$~meV.
The results of the fit are shown in Fig.~\ref{fig:exp}.

\clearpage

\clearpage

\begin{table}
\caption{
Initial adsorption, desorption, and nearest and next-nearest neighbor diffusion rates,
$R_{\rm a}$, $R_{\rm d}$, $R_{\rm nn}$, and $R_{\rm nnn}$,
respectively, for each of the four potential steps.
The rates are given in (MCSS)$^{-1}$.  The parameters are 
$k_{\rm B} T=25$~meV, $\phi_{\rm nnn}=-26$~meV, $\Delta_{\rm a}=300$~meV, $\Delta_{\rm nn}=100$~meV,
$\Delta_{\rm nnn}=200$~meV, and $\nu=1$.
}
\vspace{0.2in}
\begin{tabular}{|l|l|l|l|l|}
\hline
& \multicolumn{2}{c|}{disorder-to-order} & \multicolumn{2}{c|}{order-to-disorder} \\ \hline
& $\bar{\mu}_1=-200$~meV,& $\bar{\mu}_1=-200$~meV, & $\bar{\mu}_1=+600$~meV, & $\bar{\mu}_1=+600$~meV, \\
& $\bar{\mu}_2=+600$~meV & $\bar{\mu}_2=+250$~meV & $\bar{\mu}_2=-200$~meV & $\bar{\mu}_2=+100$~meV \\ \hline
$R_{\rm a}$ & 1 & $9.1 \times 10^{-4}$ & $2.6 \times 10^{-9}$ & $1.1 \times 10^{-6}$ \\ \hline
$R_{\rm d}$ & $3.8 \times 10^{-11}$ & $4.1 \times 10^{-8}$ & $1.4 \times 10^{-2}$ & $3.5 \times 10^{-5}$ \\ \hline
$R_{\rm nn}$ & $7.3 \times 10^{-2}$ & $7.3 \times 10^{-2}$ & $3.7 \times 10^{-2}$ & $3.7 \times 10^{-2}$ \\ \hline
$R_{\rm nnn}$ & $1.3 \times 10^{-3}$ & $1.3 \times 10^{-3}$ & $3.4 \times 10^{-4}$ & $3.4 \times 10^{-4}$ \\ \hline

\end{tabular}
\end{table}

\end{document}